\documentclass[10pt]{article} 
\usepackage[preprint]{tmlr}

\usepackage{amsmath,amsfonts,bm}

\def\eqref#1{equation~\ref{#1}}

\def\1{\bm{1}}

\DeclareMathAlphabet{\mathsfit}{\encodingdefault}{\sfdefault}{m}{sl}
\SetMathAlphabet{\mathsfit}{bold}{\encodingdefault}{\sfdefault}{bx}{n}

\definecolor{mydarkblue}{rgb}{0,0.08,0.45}
\usepackage[colorlinks,citecolor=mydarkblue,urlcolor=mydarkblue,linkcolor=mydarkblue]{hyperref}
\usepackage{url}
\usepackage{graphicx}
\usepackage{enumitem}
\usepackage{booktabs}
\usepackage{threeparttable}
\usepackage{multirow}
\usepackage{tabularx}
\usepackage{subcaption}
\usepackage{colortbl}
\title{Interpretable Machine Learning for Weather and Climate \\ Prediction: A Survey}

\author{\vspace{1.3mm}\name Ruyi Yang\hspace{0.5mm}$^{1}$ \email yangruyi853@gmail.com \\
\vspace{1.3mm}\name Jingyu Hu\hspace{0.5mm}$^{2}$ \email  ym21669@bristol.ac.uk \\
\vspace{1.3mm}\name Zihao Li\hspace{0.5mm}$^{3}$ \email  lizihao9885@gmail.com \\
\vspace{1.3mm}\name Jianli Mu\hspace{0.5mm}$^{1}$ \email  mujl668@sina.com \\
    \vspace{1.3mm}\name Tingzhao Yu\hspace{0.5mm}$^{1}$ \email  tsingzao@hotmail.com \\
    \vspace{1.3mm}\name Jiangjiang Xia\hspace{0.5mm}$^{4}$ \email  xiajj@tea.ac.cn \\
    \vspace{1.3mm}\name Xuhong Li\hspace{0.5mm}$^{5}$ \email  jacqueslixuhong@gmail.com \\
\vspace{1.3mm}\name Aritra Dasgupta\hspace{0.5mm}$^{6}$ \email  aritra.dasgupta@njit.edu \\
     \vspace{1.3mm}\name Haoyi Xiong\hspace{0.5mm}$^{5}$ \email  haoyi.xiong.fr@ieee.org \\
    \addr
    $^{1}$Public Meteorological Service Center, China Meteorological Administration \hspace{1mm} 
    $^{2}$University of Bristol \hspace{1mm} \\
    $^{3}$Zhejiang University \hspace{1mm}
    $^{4}$Institute of Atmospheric Physics, Chinese Academy of Science
    $^{5}$Baidu Inc. \hspace{1mm} \\
    $^{6}$New Jersey Institute of Technology \hspace{1mm} 
}

\def\month{MM}  
\def\year{YYYY} 
\def\openreview{\url{https://openreview.net/forum?id=XXXX}} 

\begin{document}

\maketitle

\begin{abstract}
Advanced machine learning models have recently achieved high predictive accuracy for weather and climate prediction. However, these complex models often lack inherent transparency and interpretability\footnote{In this paper, the terms "explanation" and "interpretation," as well as "explainability" and "interpretability," and "explainable" and "interpretable" are used interchangeably.}, acting as "black boxes" that impede user trust and hinder further model improvements. As such, interpretable machine learning techniques have become crucial in enhancing the credibility and utility of weather and climate modeling. In this survey, we review current interpretable machine learning approaches applied to meteorological predictions. We categorize methods into two major paradigms: 1) Post-hoc interpretability techniques that explain pre-trained models, such as perturbation-based, game theory based, and gradient-based attribution methods. 2) Designing inherently interpretable models from scratch using architectures like tree ensembles and explainable neural networks. We summarize how each technique provides insights into the predictions, uncovering novel meteorological relationships captured by machine learning. 
Lastly, we discuss research challenges around achieving deeper mechanistic interpretations aligned with physical principles, developing standardized evaluation benchmarks, integrating interpretability into iterative model development workflows, and providing explainability for large foundation models. 
\end{abstract}

\section{Introduction}
Weather and climate change have a significant impact on social, economic, and environmental systems around the world. Therefore, accurate weather forecasting and climate prediction are crucial to hazard preparation, resource management, and understanding long-term climate change. Traditionally, these predictions have relied heavily on complex numerical models that solve fundamental physics equations influencing atmospheric dynamics~\citep{richardson1922weather}, such as Numerical Weather Prediction (NWP) models and General Circulation Models (GCMs). However, these physics-based numerical predictions have some major limitations, including uncertainties in initial conditions, incomplete representations of sub-grid processes, and constraints on spatial resolution and computing power. In recent years, machine learning (ML) techniques, particularly deep learning models, have achieved dramatic progress in processing massive datasets, characterizing spatial features~\citep{du2020advances}, mining time correlations~\citep{lee2020machine}, super-resolution downscaling~\citep{leinonen2020stochastic}, and extracting spatial-temporal series in model predictions~\citep{guo2021learning}. As a result, an increasing number of scientific and business entities have incorporated machine learning into weather forecast and climate prediction~\citep{yu2023temporal,qian2023seasonal,chkeir2023nowcasting,yang2022hourly,yu2022terrain,arcomano2020machine,weyn2019can,scher2018predicting}. The outstanding performance of large foundation models of weather forecast such as ClimaX~\citep{nguyen2023climax} and GraphCast~\citep{lam2023learning} shows the potential of machine learning-based prediction models in meteorological prediction.

Despite their predictive capabilities, most advanced ML models used for meteorology are usually regarded as "black boxes", lacking inherent transparency in their underlying logic and feature attributions~\citep{du2019techniques,deng2021unified,xiong2024towards}. This lack of interpretability poses major challenges. First, it reduces trust from domain experts, such as meteorologists, who may be reluctant to rely on unexplained model outputs for high-stakes decision making. Second, it hinders further model refinement, as developers cannot easily diagnose errors or identify which relationships the models have captured. Third, opaque ML models provide limited insight into the fundamental atmospheric processes that lead to their predictions.

\begin{figure}[t]

\centering

\begin{subfigure}[b]{0.99\textwidth}
\includegraphics[width=\textwidth]{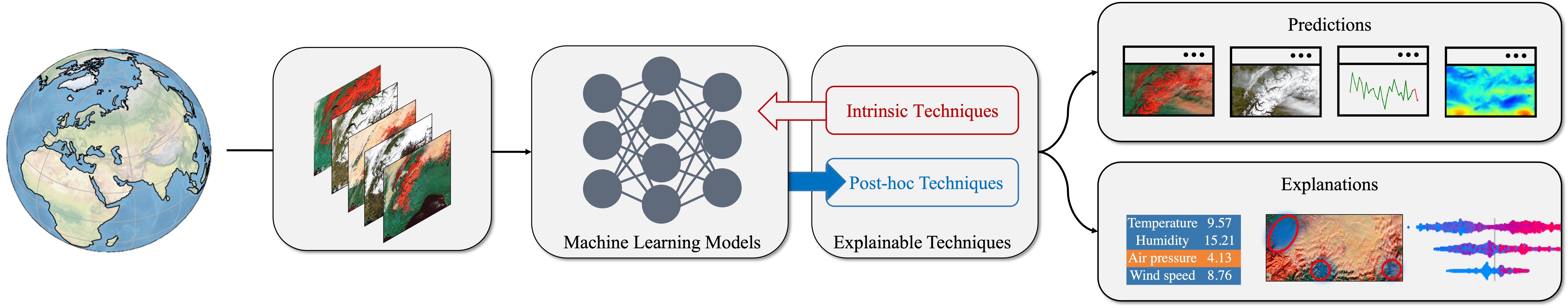}
\caption{Explainable Machine Learning in weather and climate prediction}
\label{fig:workflow-xai}
\end{subfigure}

\begin{subfigure}[b]{0.99\textwidth}
\includegraphics[width=\textwidth]{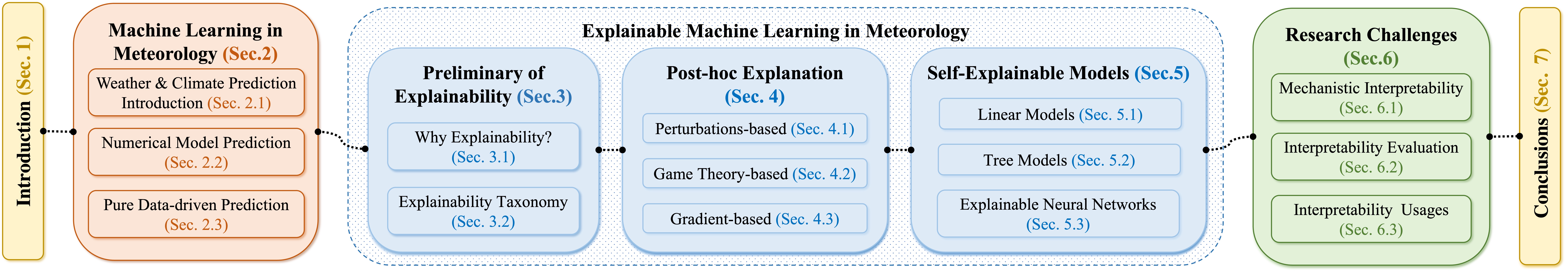}
\caption{The Structure of the Survey}
\label{fig:structure-tmp}
\end{subfigure}
\caption{Explainable machine learning in meteorological prediction and the structure of the survey}
\end{figure}

To address these limitations, explainable machine learning (Fig.~\ref{fig:workflow-xai}) techniques have become essential to enhance trust in predictions, facilitate further model improvements, and uncover new meteorological insights~\citep{labe2023changes,arrieta2020explainable,mcgovern2019making}.
Despite initial progress in applying explainability techniques in weather and climate prediction applications, a systematic framework is still needed to summarize current research and challenges. However, existing surveys either focus mainly on reviewing explainability methods for general ML fields, as exemplified by recent surveys~\citep{du2019techniques,murdoch2019interpretable}, or review how to apply ML approaches to weather and climate predictions~\citep{bochenek2022machine,ren2021deep}.

Addressing this research gap, our work presents a comprehensive survey of the current advancements in applying explainability techniques across various methodological predictions. We categorize explainability into post-hoc explanations, which provide explainations for pre-trained models, and inherently interpretable models designed from scratch. For example, we examine representative post-hoc explanation methods such as SHapley Additive exPlanations (SHAP)~\citep{lundberg2017unified}, Layer-wise Relevance Propagation (LRP)~\citep{bach2015pixel}, and Gradient-weighted Class Activation Mapping (Grad-CAM)~\citep{selvaraju2017grad} for weather and climate predictions. In addition, we also analyze the strengths and weaknesses of each explainability technique and highlight promising directions for future research.

In conclusion, our survey provides a comprehensive review of interpretable machine learning applications in weather and climate prediction (Fig.~\ref{fig:structure-tmp}). Section 2 introduces machine learning in meteorology, breaking it down into two sub-categories: numerical model prediction improvement, and pure data-driven prediction. Section 3 addresses the preliminary aspects of explainability, including the reasons why we need explainability and a taxonomy of explainability techniques. Section 4 offers an overview of post-hoc explanation methods, such as perturbation-based, gradient-based, and game theory-based approaches. Section 5 introduces self-explainable models, including linear models, tree models, and explainable neural networks. Finally, Section 6 highlights the research challenges in the field, discussing mechanistic interpretability, evaluation of interpretability, the usages of interpretability, and interpretability for large foundation models in more detail.

\section{Machine Learning in Weather and Climate Prediction}
Machine learning has made significant progress in weather and climate prediction in recent years~\citep{bochenek2022machine,kashinath2021physics,yu2021atmconvgru}. 
Weather prediction is generally defined as the forecast of various meteorological elements at a certain time for up to two weeks, whereas climate prediction operates on longer timescales, ranging from a month to decades. The latter mainly relies on simulations that incorporate a wide range of variables, including atmospheric chemistry, ocean currents, land surface processes, and ice dynamics~\citep{giorgi1991approaches}. 
In the following, we first introduce datasets for weather and climate prediction and then present the two categories of machine learning in weather and climate prediction: improvements of numerical model prediction and pure data-driven prediction.

\subsection{Weather and Climate Datasets}
As an important part of geoscience data, meteorological data generally contain five dimensions: meteorological variables (e.g., temperature, wind speed, humidity and air pressure), time, longitude, latitude, and altitude~\citep{wang2014meteoinfo,daly2006guidelines}. These data can be regarded as either multi-variable four-dimensional spatio-temporal data or multi-variable and multi-time three-dimensional spatial data~\citep{wang2019open}. The meteorological data used for weather and climate prediction are generally divided into two categories, observation data and numerical model data. Observation data are mainly obtained by measuring and determining atmospheric conditions and variation through special sensors and detection equipment, including ground meteorological data, upper-level meteorological data, radar meteorological data, satellite meteorological data, etc. The first two are usually site data, showing non-uniform distribution in space. The latter two are quasi-grid data, which are usually processed into grid data through interpolation and other methods. Numerical model data, on the other hand, are primarily obtained by solving mathematical physics equations that can describe atmospheric motion through numerical prediction models. These include forecast data~\citep{molteni1996ecmwf} and reanalysis data from multi-source historical meteorological data fusion~\citep{kalnay2018ncep,hersbach2020era5}. The model data are usually grid data, with latitude and longitude evenly distributed in space. Meteorological data, characterized by their vast variety and long time series, can be considered as big data. Due to the strong spatial correlation and time continuity of meteorological data, machine learning technology can be used to analyze the underlying characteristics of the data, in which observation data and reanalysis data are often used as ground-truth labels.

\subsection{Machine Learning for Numerical Model Prediction Improvements}
The improvement of numerical model prediction using machine learning can be further divided into three sub-categories: data assimilation of numerical model, physical process of numerical model and post-processing of numerical model prediction, as shown in Table~\ref{tab:ml-applications}.

\textbf{Data Assimilation of Numerical Model:}
Data assimilation is a critical process in numerical prediction models that involves integrating observational data into a numerical model to provide a more accurate initial state for forecasts~\citep{wang2022deep,gustafsson2018survey,anderson2009data}. This process relies on sophisticated algorithms that balance the latest observations with prior forecasts to correct inaccuracies in the model's initial conditions. Observational data can come from a myriad of sources such as satellites, sounding balloons, radar, and ground stations, and are crucial for capturing the state of the atmosphere at a given time. By continually incorporating real-time data, numerical prediction models can significantly improve their predictive accuracy, leading to more reliable weather forecasts. Since machine learning and data assimilation are both designed to extract the relationship between prediction and influential factors for minimizing the deviation of the predicted result from the ground-truth values, machine learning models can be used instead of traditional data assimilation schemes~\citep{he2022improving,wu2021fast,harter2012data}. Using the data assimilation method based on machine learning model can not only reduce the initial field error of numerical weather prediction, but also effectively improve the operation efficiency~\citep{arcucci2021deep,cintra2018data}.

\textbf{Physical Process of Numerical Model:}
The physical processes within numerical models are governed by the fundamental laws of physics, specifically the dynamics, thermodynamics, and conservation of mass and momentum. These models simulate the behavior of the atmosphere by discretizing it into a grid and solving the equations for each grid cell. Some complex evolution processes of small- and medium-scale systems cannot be described by normal grid scale, but they are essential for achieving the prediction performance. It is required to approximate these physical processes such as convection, cloud formation and radiation transfer, which is called parameterization~\citep{bauer2015quiet}. There always has a deviation between the parameterization scheme and the real physical process, which leads to the bias between the prediction and the observed values.
Due to powerful feature extraction and nonlinear fitting capabilities for big data, machine learning algorithms, especially deep learning models, can be used to improve or replace the physical parameterization process~\citep{bodini2020can,rasp2018deep}. Relevant research shows that deep learning models can dramatically improve the computational efficiency of the physical process of numerical prediction models, but there are certain limitations~\citep{seifert2020potential}. Stability and prediction performance can be improved by adding physical constraints to deep learning models.

\textbf{Post-processing of Numerical model Prediction:}
Numerical prediction models obtain the atmospheric state in the future period mainly by solving the equations of fluid mechanics and thermodynamics which depicts the atmospheric evolution over time. Due to the uncertainty of the initial field, inaccurate representation of physical or dynamic processes and the chaos of atmospheric motion, the forecast data of the numerical model often have systematic deviation from the observation. Additionally, owing to limitations of computing resources, the spatiotemporal resolution of model data cannot meet the needs of practical applications. Traditional statistical methods have limited ability to improve the accuracy and spatiotemporal resolution of numerical predictions. The use of machine learning based post-processing technology for numerical prediction can effectively improve the forecast quality. Moreover, the online learning technology of machine learning can update the model as the observation data is updated without retraining the model. Post-processing of numerical prediction models involves several steps to refine the raw model output into usable forecasts~\citep{ma2024statistical}. The first step is the numerical prediction model itself, where the raw data from the model are interpreted to predict the future states of the weather. Following this, post-processing technology is applied, which adjusts the model output to meet practical application needs. This process improves the performance of the model by aligning it more closely with the observed reality. The post-processing of the numerical model mainly includes bias correction and statistical downscaling. The bias correction is mainly used to correct the systematic deviation of the model forecast based on observation data~\citep{han2021deep,zhang2020correction,watson2019applying}. Statistical downscaling is used to translate large-scale information from the numerical model to a finer resolution that is more relevant for local forecasts~\citep{pan2019improving,sachindra2018statistical}. This can involve using historical observation data to adjust the model output to reflect local weather patterns more accurately.

\begin{table}[t]
\begin{centering}
\caption{Different stages of machine learning in weather and climate prediction.}
\begin{tabularx}{\textwidth}{m{3.5cm}X}
\toprule
\textbf{Stages} & \textbf{Applications}  \\
\toprule
\multirow{3}{=}{Pre-processing} & Data assimilation in weather predictions~\citep{wu2021fast,arcucci2021deep,cintra2018data,harter2012data}; Climate predictions: ~\citep{he2022improving}\\
\hline
\multirow{2}{=}{In-processing} & Improve physical parameterization in weather predictions ~\citep{seifert2020potential}; Climate predictions ~\citep{rasp2018deep} \\
\hline
\multirow{3}{=}{Post-processing}  & Bias correction: ~\citep{han2021deep,zhang2020correction,watson2019applying} ;  Statistical downscaling: ~\citep{pan2019improving,sachindra2018statistical}  \\
\hline
\end{tabularx}
\label{tab:ml-applications}
\end{centering}
\end{table}

\subsection{Pure Data-driven Machine Learning Prediction}
The exponential growth of high-resolution radar observation, satellite data, numerical model output, and other meteorological data provides the data foundation for machine learning, especially deep learning, and greatly promotes the development of pure data-driven weather and climate prediction.
Data-driven weather forecasts represent a paradigm shift from traditional numerical model-based predictions, leveraging the power of machine learning and big data analytics. 
This paradigm uses historical meteorological data to train machine learning algorithms that can identify patterns and predict future atmospheric conditions. By analyzing large datasets that include past weather events, data-driven models can uncover complex relationships and dependencies that traditional methods might overlook. These models can provide valuable insights, especially in situations where physical models struggle due to chaotic atmospheric behavior. Furthermore, the integration of machine learning techniques can enhance the speed and efficiency of forecast generation, making it possible to provide real-time updates with increased accuracy. According to the difference in driving data, it can be divided into observational data-driven prediction and numerical output-driven prediction.

\textbf{Observational Data-driven Prediction:}
Observational data-driven prediction means that all input data come from observational data, and this type of prediction model does not rely on numerical models at all. One of the most common machine learning models driven by observational data is nowcasting (0–2h)~\citep{chkeir2023nowcasting,ayzel2020rainnet,foresti2019using,agrawal2019machine,shi2015convolutional}. Due to the fact that numerical model takes several hours to reach equilibrium state and has limited forecasting ability for mesoscale systems, there exist some errors between the model's nowcasting and observations. Moreover, traditional extrapolation methods based on historical observation data (e.g., radar echoes, satellite images) such as the optical flow and cross-correlation algorithm cannot effectively predict the generation and extinction of convective system. Therefore, machine learning methods are widely used in nowcasting because they show great potential in quickly fusing a large number of observational data and effectively extracting nonlinear features~\citep{zhang2023skilful}. By transforming the nowcasting into a spatiotemporal series prediction problem, machine learning model can effectively predict the generation, evolution, and extinction of convective systems in central-eastern and southern China based on multi-source observation data including satellite infrared images, radar reflectivity, and lightning density~\citep{zhou2020deep}. 

\textbf{Numerical Output-driven Prediction:}
Numerical output-driven prediction is defined as a forecast in which the input data are partially or completely derived from the numerical model output. For prediction with longer forecast time than nowcasting, in addition to using observation data such as radar and satellite data, it is necessary to provide atmospheric circulation fields from the ground to upper-air. Specially, from short-to-medium-term weather forecast to decadal climate prediction, input factors are usually derived from numerical model reanalysis datasets~\citep{price2023gencast,chen2023fengwu,bi2023accurate,weyn2020improving}.
These predictors are analogous to the initial fields of numerical weather prediction from which machine learning models can extract dynamical features. Currently, large foundation models of meteorological forecast based on artificial intelligence released globally, such as Fuxi from Fudan University~\citep{chen2023fuxi}, GraphCast developed by Google DeepMind~\citep{lam2023learning}, and FourCastNet developed by Nvidia~\citep{pathak2022fourcastnet}, are driven by atmospheric reanalysis data in numerical models. This type of model does not achieve true end-to-end forecasting, and the quality of its results depends to some extent on the results of numerical models. In addition, although the performance of this type of prediction model for general weather is comparable to that of state-of-the-art numerical models, the forecast ability for extreme weather can be further improved. To solve the problem, a feasible method is to incorporate physical mechanisms when building machine learning based prediction models~\citep{kochkov2023neural}.

\section{Preliminary of Explainability}
In this section, we discuss the preliminary of explainability, including why do we need explainability and the explainability definition and taxonomy.

\subsection{Why Explainability?}
Despite the promising progress of machine learning in various meteorology prediction tasks, most of these complex models act as ``black boxes''. Their inherent lack of transparency and interpretability reduces trust from end users, such as meteorologists and forecasters, who remain unsure how forecasts are made. It also hampers model debugging and diagnosis by developers seeking to further improve predictive performance.
Explainability is crucial in meteorological modeling for several key reasons:

\begin{itemize}[leftmargin=*]\setlength\itemsep{-0.3em}
\item \textbf{Increase trust from end users:} Complex machine learning models such as deep neural networks can achieve high accuracy but act as ``black boxes'', lacking transparency in their predictions. This makes meteorologists reluctant to fully trust and use the outputs. Interpretable machine learning can well visualize the decision-making principles of the model, which help end users understand the model logic. If the model logic is consistent with the domain knowledge of atmospheric science, it can increase the trust of users.

\item \textbf{Help developers diagnose and improve models:} Interpretability methods highlight which meteorological factors and relationships drive predictions. This allows model developers to debug errors, check if meaningful patterns are learned, and improve model architecture and data pre-processing. The ultimate goal is to increase performance metrics such as accuracy and reliability.

\item \textbf{Gain meteorological insights:} Explaining model behavior can reveal new discoveries about weather and climate prediction mechanism. For example, interpretable machine learning can be used to uncover the impact of global heating on North Atlantic circulation~\citep{sonnewald2021revealing}. These insights further advance scientific knowledge in meteorology.

\item \textbf{Integration of data-driven ML models with physical mechanism:} 
Understanding how machine learning models make predictions can help visualize whether the logic of a weather prediction model is consistent with the physical mechanisms of meteorology. It helps identify complementary strengths between statistical machine learning methods and methods based on physical mechanisms. This ultimately enables more efficient model fusion or ensembling of these two paradigms.

\end{itemize}

In summary, interpretable machine learning is key for the adoption of ML in meteorology by increasing user trust, helping model improvement, extracting new findings, and enabling the integration of machine learning models with physical mechanism. Tailored interpretability methodologies are needed to provide meaningful explanations that meet the needs of domain experts. 

\subsection{Explainability Definition and Taxonomy}

\textbf{Definition:} Given the dataset ${D}$ comprising pairs of input values $X$ (e.g., relative humidity, wind speed, cloud amount, etc.) and the corresponding target outcomes $Y$ (e.g., horizontal visibility), the model $f: \mathcal{X} \rightarrow \mathcal{Y}$ is used to predict outcomes based on input data $\mathcal{X} \in X$. For a specific test sample $x_{i} \in D$, the explanation to the model prediction $f(x_{i})$ is represented as $g(f(x_{i}))$. It aims to clarify the reasoning behind the predictions made by the model. 

\textbf{Taxonomy:} Based on the above definition, we have the following taxonomy of explainability.
\begin{itemize}
    \item Regarding explanation contents, they are called local explanations when explaining a specific test data point $(x_{i}, y_{i})$, and global explanations when referring to the entire dataset $D$.
    \item Regarding explanation method designs, explanations are categorized as model-specific and model-agnostic. Model-specific methods are especially designed for specific model(s), while model-agnostic methods can be applied to any machine learning model \citep{ribeiro2016model}.
    \item Regarding explanation generation stages, explanations can be either post-hoc or intrinsic. Post-hoc explanations occur after $f$ is trained, while intrinsic explanations imply that $f$ is self-interpretable, explaining its predictions \citep{kakkad2023survey}.
\end{itemize}

In the following sections, we categorize the explainability analysis techniques into two major types: post-hoc explanation methods presented in Section~\ref{sec:post-hoc-explanation} and the inherently interpretable model design covered in Section~\ref{sec:Design Intrinsic Self-Explainable Model}. The former explains black-box models after they have already been trained, while the latter involves building interpretable prediction models.

\section{Post-hoc Explanation}\label{sec:post-hoc-explanation}

\begin{figure}[t]
\centering
\begin{subfigure}[b]{0.99\textwidth}
\centering
\includegraphics[width=0.95\textwidth]{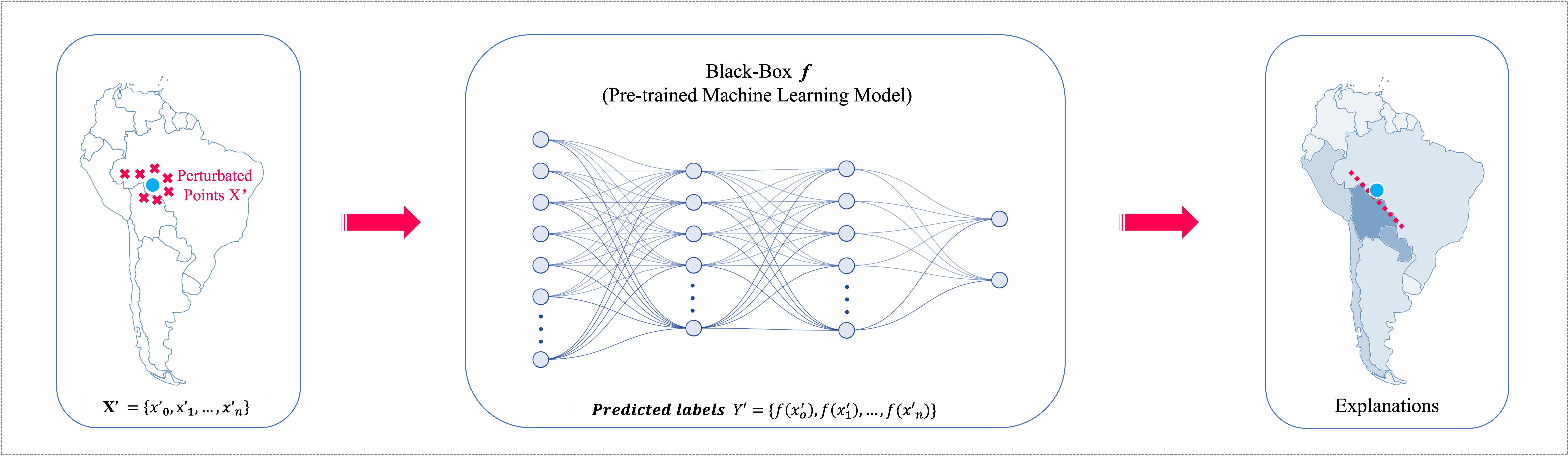}
  \caption{Perturbation-based methods.}
  \label{fig:perturbation_xai}
  \vspace{2pt}
\end{subfigure}
\begin{subfigure}[b]{0.99\textwidth}
\centering
  \includegraphics[width=0.95\textwidth]{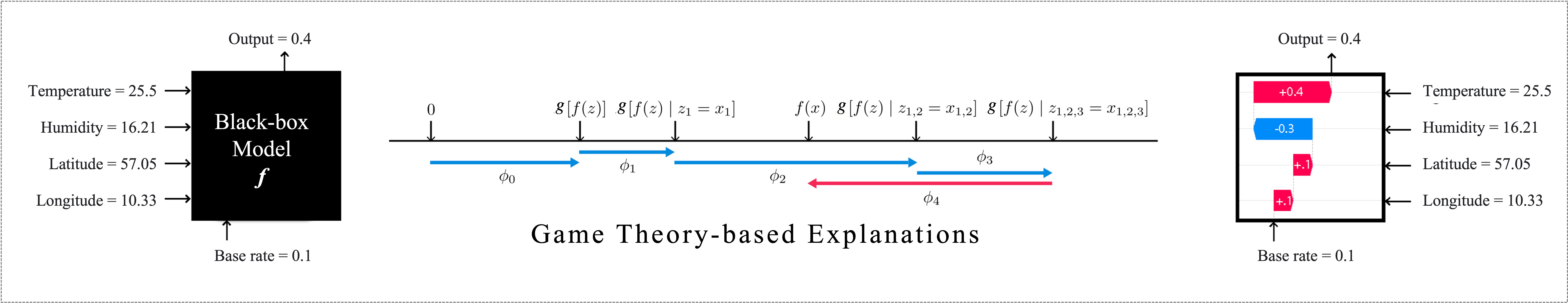}
  \caption{Game Theory-based methods.}
  \label{fig:game_xai}
\end{subfigure}
\begin{subfigure}[b]{0.99\textwidth}
\centering
  \includegraphics[width=0.95\textwidth]{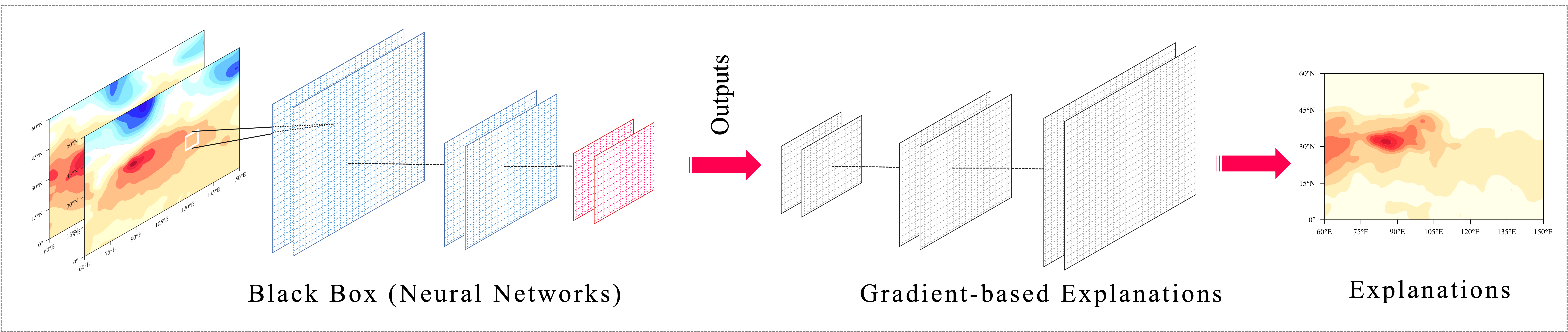}
  \caption{Gradient-based methods.}
  \label{fig:gradiet_xai}
\end{subfigure}
\caption{Three major families of post-hoc explanation methods in weather and climate prediction: (a) Perturbation-based explanation, (b) Game Theory-based explanation (Referred \& modified from SHAP \citep{lundberg2017unified}), (c) Gradient-based explanation.}
\label{fig:post-hoc}
\end{figure}

Post-hoc explanation techniques aim to interpret the predictions of machine learning models after they have already been trained, without changing the underlying models themselves~\citep{deng2024unifying,fu2021differentiated,wang2020score}. 
In the following, we categorize the local explanation methods into three main categories: perturbation-based methods, game theory based methods, gradient-based methods (see Fig.~\ref{fig:post-hoc}). The prediction tasks of these three categories of methods can be found in Table~\ref{table:task_xai}. Besides, we also briefly introduce explainability methods that cannot be categorized into these three categories. We summarize how these post-hoc attribution methods can be used to explain weather and climate ML predictions. 

\begin{table}[t]
\caption{An overview of three categories of post-hoc explanation methods: Perturbation-based, Game Theory-based and Gradient-based methods. }
\begin{tabularx}{\textwidth}{m{3.4cm}m{4.8cm}X}
\hline
\textbf{Methods} & \textbf{Applcations} & \textbf{Prediction Tasks} \\
\hline
\rowcolor[HTML]{EFEFEF}\multicolumn{3}{c}{Perturbation Based Post-hoc Explanation} \\ 
\multirow{4}{=}{Local Interpretable Model-agnostic Explanations (LIME)} & \citep{valdes2021machine} & Water vapor patterns prediction \\
 & \citep{gibson2021training} & Seasonal precipitation prediction \\
 & \citep{rajasekaran2023hybrid} & Irradiance, temperature and wind speed prediction \\ \hline
 & \citep{rasp2018neural} & Ensemble weather forecasts \\
 & \citep{molina2021benchmark} & Severe convective storms classification\\
 & \citep{ghada2022stratiform} & Stratiform and convective rain classification \\
\multirow{-4}{=}{Permutation Feature Importance (PFI)} & \citep{shield2022diagnosing} & Supercell thunderstorms prediction \\
\hline
\rowcolor[HTML]{EFEFEF} \multicolumn{3}{c}{Game Theory Based Post-hoc Explanation}\\ 
\multirow{3}{=}{Shapley Value} & \citep{thanh2021interpretable} & Horizontal visibility predictions \\
 & \citep{leinonen2023thunderstorm} & Thunderstorm nowcasting \\
 & \citep{wang2023deep} & Tropical cyclone wind radii predidction \\ \hline
 & \citep{lu2021regional} & Heavy precipitation prediction \\
 & \citep{gensini2021machine} & Severe weather events prediction \\
 & \citep{dutta2022interpretation} & Short-term premonsoon thunderstorms prediction \\
 & \citep{griffin2022predicting} & Rapid intensification of tropical cyclones prediction\\
\multirow{-6}{=}{SHapley Additive exPlanations (SHAP)} & \citep{silva2022using} & Earth system model errors prediction \\
\hline
\rowcolor[HTML]{EFEFEF} \multicolumn{3}{c}{Gradient Based Post-hoc Explanation} \\ 
\multirow{4}{=}{Gradient-weighted Class-Activation Mapping (Grad-CAM)} & ~\citep{higa2021domain} & Typhoon intensity classification \\
 & ~\citep{rampal2022high} & Daily rainfall prediction\\ 
 & ~\citep{renault2023sar} & Precipitation nowcasting \\
 & ~\citep{reulen2024ga} & Extreme precipitation nowcasting \\ \hline
 & ~\citep{espeholt2022deep} & Precipitation prediction \\
 & ~\citep{gonzalez2023using} & Daily temperature downscaling \\
\multirow{-3}{=}{Integrated Gradients (IG)} & ~\citep{hu2023deep} & Probabilistic precipitation forecast \\ \hline
 & ~\citep{hilburn2020development} & Synthetic radar reflectivity estimation \\
 & ~\citep{barnes2020indicator} & Annual-mean temperature and precipitation prediction \\
 & ~\citep{toms2020physically} & Surface temperature anomalies prediction and ENSO phase identification \\
 & ~\citep{toms2021assessing} & Surface temperature anomalies prediction \\
 & ~\citep{toms2021testing} & Madden-Julian Oscillation phase identification \\
 & ~\citep{sonnewald2021revealing} &  Tracking global heating with ocean regimes\\
 & ~\citep{zhuo2021physics} & Tropical cyclone intensity and size estimation\\
 & ~\citep{retsch2022identifying} & Convective area and organization prediction\\
 & ~\citep{legler2022combining} & Convective-scale model parameters estimation \\
 & ~\citep{li2023probabilistic} & Probabilistic convective initiation nowcasting\\
\multirow{-11}{=}{Layer-wise Relevance Propagation (LRP)} & ~\citep{liu2023deep} & Short-term station precipitation prediction \\
\bottomrule
\label{table:task_xai}
\end{tabularx}
\end{table}

\subsection{Perturbation-based Forward Propagation Methods}
Perturbation techniques systematically alter input features to quantify the effect on predictions (Fig.~\ref{fig:perturbation_xai}). For example, methods like Local Interpretable Model-Agnostic Explanations (LIME)~\citep{ribeiro2016should} randomly generate perturbed inputs. The prediction difference when removing or perturbing a feature highlights its importance. This type of model can be used to explain the predictions of any machine learning model. In the following, we focus on two representative perturbation-based methods, including LIME and permutation feature importance.

\subsubsection{LIME Explanation}

The LIME method mainly performs small perturbations on target input features within the local linear neighborhood and then uses a simple linear model to visualize important features to achieve the purpose of explanation~\citep{ribeiro2016should}. The method is easy to understand and can be applied to various machine learning models. For example, one study employs the LIME method to reflect the behavior of the random forest classifier in the water vapor patterns detected by meteorological satellites ~\citep{valdes2021machine}. In particular, in 2013 and 2015, the importance of certain masks used in the model changes significantly, with masks 11 and 27 being the most important in both years, but with their ranks reversed in 2015. Additionally, some masks disappear from the top 15 subset when moving from 2013 to 2015, and others not previously in the top 15 appear in 2015, indicating significant temporal variations in the contributing factors to the model's predictions. Another study applies LIME to interpret the irradiance, temperature and wind speed predictions made by SRNN-LSTM hybrid models~\citep{rajasekaran2023hybrid}. The research highlights the significance of meteorological feature correlations and time steps towards model outputs, which are analyzed using LIME frameworks. It is concluded that the explanations generated by LIME comply with the theoretical constructs of the features involved in the model. Since this method works on a single prediction and can only be used for local explanation, it is often used in conjunction with other explainability methods to improve a more comprehensive explanation. For example, the LIME method, together with other explanation methods like partial dependence plots, is used to interpret seasonal precipitation prediction made by a random forest model for the western United States~\citep{gibson2021training}. Specifically, the LIME method focuses on explaining the model's decision-making process for individual seasonal forecasts, and the latter are used to make global interpretation. Using LIME, the study analyzes two case studies: one where the model correctly predicts the seasonal cluster in 2005, and another where it incorrectly predicts the seasonal cluster in 2016. In the 2016 case, weak El Niño conditions largely contribute to this incorrect prediction. This analysis demonstrates the capacity of LIME to provide insights into the specific factors that influence individual weather forecasts, thus enhancing the interpretability of complex weather prediction models. 

In this survey, we showcase the performance of LIME by using it to interpret and visualize an MLP-based temperature prediction task. The importance of input features in two cases is shown in Fig.~\ref{fig:sub1} and~\ref{fig:sub2}.

\subsubsection{Permutation Feature Importance}

The Permutation Feature Importance (PFI) method is first proposed to measure the importance of input variables in random forests~\citep{breiman2001random}. The input variables are ranked based on the influence of random perturbations on prediction errors, with larger difference in skill indicating greater importance~\citep{gagne2019interpretable}. 
In one study, the PFI is applied to determine the relative importance of different features in ensemble weather forecasts~\citep{rasp2018neural}. This method involves randomly shuffling each predictor or feature in the validation set one at a time and observing the increase in mean Continuous Ranked Probability Score (CRPS) compared to the unpermuted features. The results of this interpretation method show that it is necessary to extract local information to improve the accuracy of probabilistic temperature forecast.
Another study employs the PFI method to analyze the impact of specific meteorological variables on the performance of a Convolutional Neural Network (CNN) in classifying thunderstorms~\citep{molina2021benchmark}. This process involves 500 permutations for each of the 20 variables, allowing for a comprehensive assessment of variable importance in the context of thunderstorm classification. 
In addition, PFI is utilized to rank input features based on their importance for the performance of machine learning models in classifying rain types~\citep{ghada2022stratiform}. This approach permutes a specific feature to disrupt its association with the target variable, followed by predictions using the modified dataset. The resulting change in model performance, particularly in the Area Under the Curve (AUC), indicates the significance of the permuted feature for the model's predictive performance. The reflectivity at the lowest layer and the average pectral width in the layers below separation level are most influential in the model's predictions for classifying rain types.

This PFI approach mentioned above helps to identify which features significantly impact the model's performance without the need to re-estimate the model for each feature omitted. However, it is important to note that this method does not capture colinearities between features.
To fully account for the correlation between input variables, the multipass permutation variable importance method~\citep{lakshmanan2015polarimetric} is developed to interpret machine learning models for forecasting supercell thunderstorms~\citep{shield2022diagnosing}. This method involves initially assessing the model's performance on a test dataset, then permuting a single input variable and re-evaluating the model's performance with the altered data. The variable whose permutation results in the most significant decrease in performance is considered the most important. This procedure is run repeatedly, keeping previously ranked variables permuted, until all variables are ranked. This approach, while more computationally intensive, provides a more precise understanding of the variables' impact on the model, especially in the presence of correlations between variables.

\subsection{Game Theory based Methods}
Game theory-based explanation methods like Shapley values and SHapley Additive exPlanations (SHAP) explain a model's predictions by attributing impact to input features (Fig.~\ref{fig:game_xai}). They do this by comparing a prediction to what it would be without each feature. These techniques can generate consistent, model-agnostic explanations based on game theory. However, they scale poorly computationally for complex models with many features. In the following, we introduce how both methods can be used to explain the meteorological predictions made by ML models.

\subsubsection{Shapley Value Explanation}
The Shapley values method from cooperative game theory quantifies the contribution of each feature to a model's output by comparing what the prediction would be with and without that feature. This method calculates the marginal contribution of a feature by iterating through all possible combinations, or coalitions, of features.
For instance, the Shapley value method is utilized for local interpretation of specific weather predictions~\citep{thanh2021interpretable}. This method assesses the contribution of each meteorological feature to the difference between the actual prediction and the average prediction. In a case study of horizontal visibility in Kemi, Finland, air temperature is found to have the most positive contribution, while cloud amount had the most negative impact. Besides, the Shapley value method is used to assess the importance of each data source in predicting thunderstorm hazards, enhancing prediction explainability~\citep{leinonen2023thunderstorm}. This approach calculates the contribution of each predictor to improving a performance metric, employing game theory for fair interpretation. The Shapley values represent the weighted average of improvements achieved by adding a predictor to a set that previously did not contain it. This method requires computing the metric for each subset of predictors, including the empty set, which is feasible in this study due to the limited number of data sources. The sum of all Shapley values is equal to the total improvement provided by all predictors. The results demonstrate that the weather radar data play the most important role in prediction of the three types of disasters: lightning, hail and heavy precipitation.
Furthermore, the Shapley value method is employed to interpret deep learning predictions of tropical cyclone wind radii~\citep{wang2023deep}. This approach is used to generate heat maps, highlighting features with higher relevance to the tropical cyclone wind radius predictions. Larger values in these maps indicate features of greater significance to the prediction outcomes, providing insight into the individual contributions of the variables in the model.

\begin{figure}[t]
    \centering
    \begin{subfigure}[b]{0.45\textwidth}
    \includegraphics[width=0.8\textwidth]{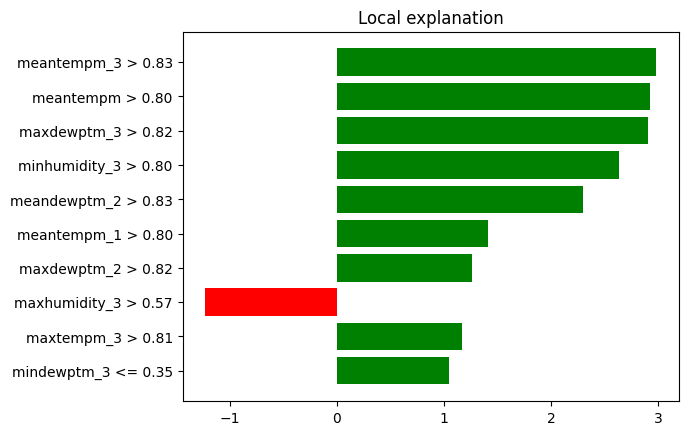}
    \caption{LIME output for 2017/7/5}
    \label{fig:sub1}
    \end{subfigure}
     \begin{subfigure}[b]{0.45\textwidth}
    \includegraphics[width=0.8\textwidth]{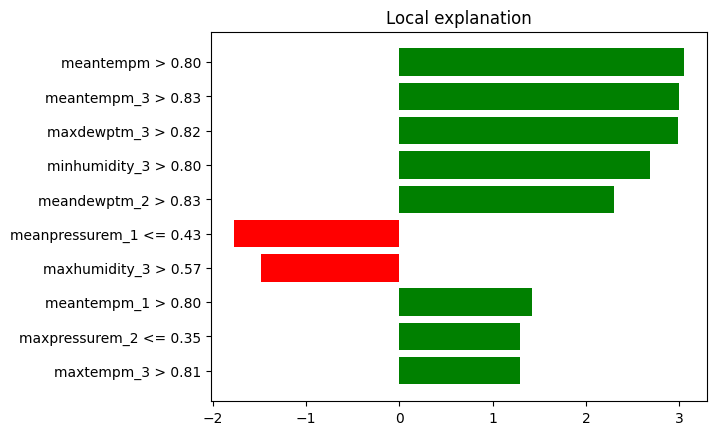}
    \caption{LIME output for 2017/7/26}
    \label{fig:sub2}
    \end{subfigure}

    \begin{subfigure}[b]{0.45\textwidth}
    \includegraphics[width=0.8\textwidth]{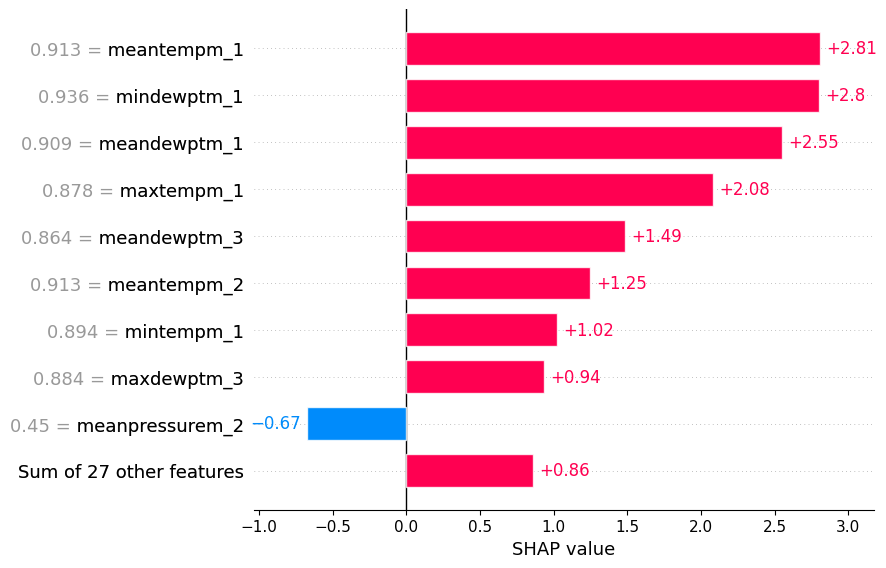}
    \caption{SHAP output for 2017/7/5}
    \label{fig:sub3}
    \end{subfigure}
     \begin{subfigure}[b]{0.45\textwidth}
    \includegraphics[width=0.8\textwidth]{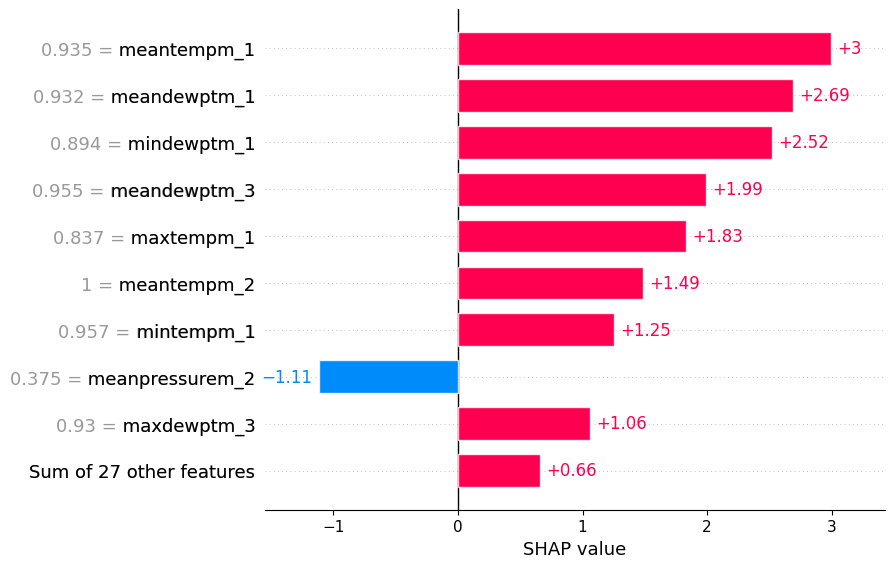}
    \caption{SHAP output for 2017/7/26}
    \label{fig:sub4}
    \end{subfigure}

    \caption{ The feature attribution results of different input features in MLP prediction models using LIME and SHAP interpretation methods. The MLP model is used to forecast the weather temperature task. The data comes from the weather data of Lincoln, Nebraska, which is provided by Weather Undergroun and covers 997 days since January 4, 2015~(Here, MLP model and data are both from \url{https://github.com/Rite188/Simple-DNN-on-weather-forcast}).}
\end{figure}

\subsubsection{SHAP Explanation}
Although Shapley value interpretation can consider all feature coalitions comprehensively, directly computing Shapley values for all $2^n$ feature subsets is computationally expensive for high dimensional models. SHAP is built on this framework to efficiently estimate Shapley values through conditional expectations~\citep{lundberg2017unified}. This enables model explanations to scale to higher dimensions by approximating Shapley values, simplifying assumptions to enable faster computation. Due to the good trade-off between accuracy and feasibility, the method is widely used in weather and climate predictions.
For example, the SHAP method is used to interpret the contributions of meteorological features in a deep learning model for forecasting heavy precipitation~\citep{lu2021regional}. For visual analysis, the SHAP values of the 20 most important features such as thermodynamic and dynamic parameters are analyzed using 500 minority samples. It offers insights into both the relative importance of convection parameters and their positive or negative contributions to the heavy precipitation predictions. Another study utilizes SHAP to interpret the output of machine learning models predicting severe weather events~\citep{gensini2021machine}. This method allows for the identification of thresholds where each predictor variable begins to notably impact the probabilistic contribution to the forecast outcome in severe or significant-severe weather classification, providing a deeper understanding of the model's predictions for tornadoes and hail.
Additionally, SHAP is applied to quantify the contributions of different predictor variables in machine learning models for short-term predictions of pre-monsoon thunderstorms~\citep{dutta2022interpretation}. 
The study uses SHAP to determine the relevance of different predictor variables, thereby interpreting the models' decisions toward thunderstorm prediction and validating the models with domain knowledge.
Besides, the SHAP method is utilized to assess the impact of individual features on the probability of rapid intensification in tropical cyclones~\citep{griffin2022predicting}. 
The sum of SHAP values for all input features is found to be slightly lower than the rapid intensification probability, a difference that grows with increasing rapid intensification probability. This methodology offers a clear and interpretable understanding of how individual features influence the model's performance in predicting rapid intensification in tropical cyclones. Lastly, SHAP values are used to analyze the contribution of input variables to earth system model errors predictions made by XGBoost model~\citep{silva2022using}. Specifically, the SHAP framework evaluates the sum of contributions from each input feature and the average predicted value to understand their impact on lightning predictions errors. The result demonstrates the errors in lightning prediction are highly correlated with the effects of convective processes and surface heterogeneity, highlighting the potential of SHAP in characterizing and exploring errors in Earth system models.

In this survey, we also showcase the explanation performance of SHAP by providing the results of two cases using it to explain the importance of input variables, as shown in Fig.~\ref{fig:sub3} and~\ref{fig:sub4}.

\subsection{Gradient-based Backpropagation Methods}

Gradient-based feature attribution explains a prediction by analyzing input feature derivatives with model output (Fig.~\ref{fig:gradiet_xai}). Saliency methods like Gradient-weighted class-activation mapping (Grad-CAM)~\citep{selvaraju2017grad}, Integrated Gradients (IG) ~\citep{sundararajan2017axiomatic} and Layer-wise Relevance Propagation (LRP) determine influential input variables by computing gradients. Some representative gradient-based interpretation results are shown in Fig. 4. In the following, we introduce how these gradient-based attribution methods can be used to explain weather and climate predictions made by machine learning models.

\subsubsection{Grad-CAM Explanation}

Grad-CAM obtains the weight of each channel by calculating the gradient to identify the areas that influence the decision-making basis. We can use this method to visualize any layer without changing the model structure. Current research mainly applies this method to various CNN-based deep learning models.
For example, one recent study implements Grad-CAM to identify spatial locations in input fields that significantly support daily rainfall predictions made by a CNN for a specific output location~\citep{rampal2022high}. 
Grad-CAM does not specify which individual predictor variable had the strongest influence, but instead shows the importance of spatial locations in the aggregated predictor space. This approach aids in making the model's predictions more transparent and trustworthy, thereby potentially leading to new insights in complex systems. 
Another study discusses the integration of meteorological domain knowledge into deep learning models for improving typhoon intensity classification from satellite images~\citep{higa2021domain}. Specifically, it highlights the use of Grad-CAM to visualize which areas of the satellite images are most important for predicting typhoon intensity. By preprocessing images with fisheye distortion to emphasize the typhoon's eye and surrounding cloud distributions, the model achieved higher classification accuracy. Grad-CAM visualizations confirmed that the model focuses on meteorologically significant regions for intensity classification, aligning with expert domain knowledge.

Grad-CAM can also be utilized for global analysis via analyzing the importance and interaction of different layers in the machine learning model. For example, one study proposes a novel UNet-based architecture for precipitation and cloud cover nowcasting tasks~\citep{renault2023sar}. 
To provide comprehensive explanations, they apply Grad-CAM to visualize activation heatmaps at different layers in the encoder and decoder. Analyzing these heatmaps reveals how the residual connections and depthwise separable convolutions interact, with the two paths switching importance between encoder and decoder. The heatmaps also show how the model progresses from detecting borders to focusing on cloud centers. For example, deeper network levels show activations in more abstract areas, with significant activations in select spots. Overall, using Grad-CAM on multiple layers provides global and layered explanations that increase understanding of the model's precipitation nowcasts.
Furthermore, the same multiple layer analysis explanation framework has been applied to explain predictions made by a novel generative adversarial network designed for extreme precipitation nowcasting task~\citep{reulen2024ga}. There are some interesting findings. For example, the precipitation map encoder shows higher activation in areas of high precipitation at shallow depths, while the precipitation mask encoder activation maps correlate with input precipitation at first and predicted precipitation at the next depth. At deeper levels, the activation areas become more abstract for both encoders. Overall, the encoder activation heatmaps become less directly linked to precipitation levels as depth increases.

\subsubsection{Integrated Gradients Explanation}
In addition to Grad-CAM, IG is another representative gradient-based explainability technique.
It measures the impact of input changes on model predictions in deep learning, particularly in complex models such as U-Net. This technique is known for its ability to overcome issues like gradient saturation, common in standard gradient-based methods~\citep{sundararajan2017axiomatic}. By integrating the product of input values and gradients, IG assesses the sensitivity of outputs to inputs, using a blurred input as a baseline for comparison. Since this method does not alter the model's architecture and is effective for visualizing how input perturbations affect forecasts, it is widely used in weather forecast. For example, IG has been used to understand MetNet-2 learning process, a physically independent probabilistic weather model based on deep neural networks, shedding light on the interaction between various meteorological variables in precipitation forecast~\citep{espeholt2022deep}. IG attributes the network's precipitation predictions to specific input variables, revealing that while the absolute vorticity's influence is minimal for near-term forecasts, it becomes more significant for predictions up to 12 hours ahead. This aligns with the geostrophic theory, where positive vorticity at higher altitudes correlates with upward motion lower in the atmosphere, setting the stage for potential convection, a precursor to precipitation. The IG-based analysis in another study indicates that precipitation intensity and uncertainty are highly responsive to input precipitation changes, with integrated vapor transport and integrated water vapor also identified as significant factors in U-net models for precipitation prediction~\citep{hu2023deep}. 
For climate prediction, one study uses IG to integrate gradients along a path from the input to a baseline, offering a more accurate representation of feature importance in climate-related CNN applications~\citep{gonzalez2023using}. They quantify the influence of input variables on a temperature downscaling, helping to identify critical predictor variables and their spatial regions of influence.

\begin{figure}[t]
  \centering
  \includegraphics[width=0.85\textwidth]{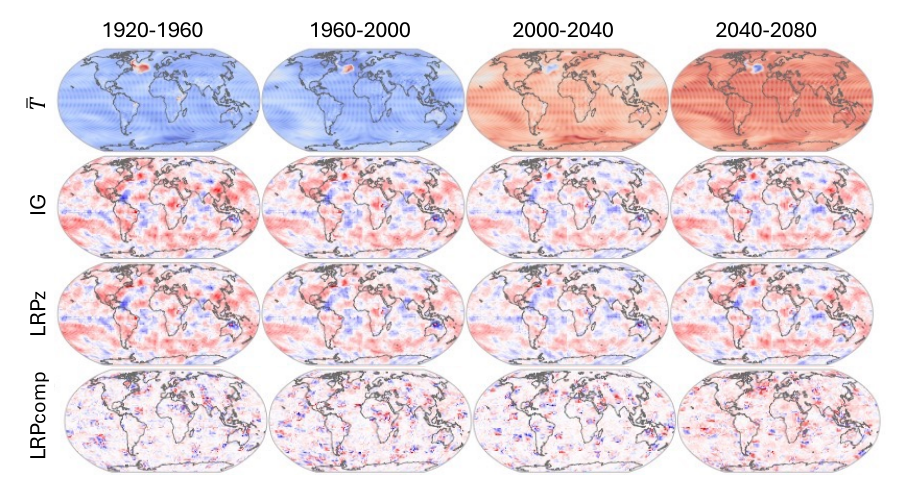} 
  \caption{ Visualization comparison between three explainability methods. Here $\text{LRP}_{comp}$ and $\text{LRP}_{z}$ use two different rules of backpropagation. The figure is adapted from Bommer et al.~\citep{bommer2023finding}.
  }

\end{figure}

\subsubsection{Layer-by-layer Backpropagation}
Deep neural networks are often extremely complex, while each layer of the neural network is relatively simple (e.g., deep features are usually a linear summation of shallow features and nonlinear activation function). This facilitates the analysis of the importance of shallow features for deep features. Therefore, this type of algorithm estimates the importance of intermediate features and propagates these importance values layer by layer to the input layer, to determine the importance of the input units. This algorithm includes Layer-wise Relevance Propagation (LRP)~\citep{bach2015pixel,ebert2020evaluation}, deep Taylor decomposition~\citep{montavon2017explaining}, etc. The fundamental difference between these backpropagation algorithms lies in the different rules they adopt for layer-by-layer propagation of importance. In the following, we will introduce the application of LRP in weather forecast and climate predictions.

For weather forecast, LRP is often used to account for predictors that have a significant impact on deep learning models. For example, one study employs LRP to investigate the interaction between deep convection in the tropics and the large-scale atmosphere when using Multi-Layer Perceptrons (MLP)\footnote{Due to historical reasons, multilayer perceptrons (MLPs) are also referred to as artificial neural networks (ANNs) or feedforward neural networks (FNNs) in the literature. For consistency, this paper uses the term MLP.} to predict convective area and organization~\citep{retsch2022identifying}. It finds that large-scale vertical velocity significantly impacts both convective area and organization, with its influence more pronounced in predicting convective area. While thermodynamic factors like atmospheric moisture affect convective area predictions, they are less important for convective organization, where horizontal wind fields play a more significant role. Another article addresses the challenge of estimating parameters not well-defined physically in convection-permitting numerical weather prediction models, particularly for cloud
representation~\citep{legler2022combining}. It uses Bayesian Neural Networks (BNNs) and Bayesian approximations of point estimate MLP, to predict several parameters of a modified shallow-water model based on atmospheric state observations or analysis. The study utilizes LRP to understand how these neural networks learn, revealing that the networks selectively focus on a few grid points characterized by strong winds and rain for making parameter predictions. Additionally, one article applies the LRP method to a Station-based Precipitation Post-processing Model (SPPM) designed to enhance the accuracy of medium-range station precipitation forecast with deep-learning algorithms~\citep{liu2023deep}. Specifically, the LRP method is used to assess the sensitivity of the predictors, revealing that total precipitation from the NWP is the most crucial and sensitive factor, particularly for larger forecast grades. The importance of low-level (850 hPa) field, single-level field, and geographic variables is also examined, which is in agreement with the meteorological domain knowledge.

Not only can we use LRP to estimate the importance of predictors, we can also use it to show where the neural network focuses primarily when making predictive decisions. To visualize the underlying logic of a CNN-based deep learning model decisions when assimilating GOES-R series observations in precipitating scenes, a study employs LRP as an attribution method, demonstrating the CNN's synergistic use of radiance and lightning information~\citep{hilburn2020development}. Lightning data particularly help in identifying important neighboring locations. The study investigates the sensitivity to radiance gradients, indicating that sharper gradients elicit stronger responses in predicted radar reflectivity, and highlights the unique value of lightning observations in pinpointing locations of strong radar echoes. This research also explored the application of LRP to regression tasks. Besides, DeepTCNet, a CNN-based model for estimating tropical cyclone intensity and wind radii, uses LRP to interpret its decision-making process~\citep{zhuo2021physics}. LRP attributes which input features most affect the model's output, overcoming the limitations of saliency maps by considering the redistribution of prediction values across input features, thereby providing a more coherent explanation for the model's predictions. This interpretative approach is crucial in enhancing trust in DeepTCNet's forecasts for tropical cyclone characteristics and has the potential to explore more complex tropical cyclone mechanisms.
In addition, LRP can also be applied to more complex deep learning models. The CIUnet model, leveraging U-net architecture and Himawari-8 data, effectively forecasts convective initiation with a high detection rate and low false alarms~\citep{li2023probabilistic}. LRP analysis validates the model's accuracy in pinpointing crucial regions and features for precise convective initiation predictions. Key input factors include brightness temperature differences between spectral channels and terrain height.

In addition to weather forecast, there is also some research focus on the application of LRP in climate prediction. In particular, a study explores the use of LRP in MLP for seasonal surface temperature anomalies prediction and El Niño Southern Oscillation (ENSO) phase identification~\citep{toms2020physically}. LRP helps explain neural network decisions by identifying the relevance of each input feature for the network's output on a sample-by-sample basis, effectively creating a heatmap of relevance. Specifically, LRP is applied to study El Niño events, with the relevance values from LRP for each sample normalized to a range of 0 to 1. This normalization ensures equal weighting of relevances across samples when composing the overall relevance heatmap. The study demonstrates that LRP can trace the reasoning of a neural network's decisions, highlighting its potential to reveal meaningful geoscientific insights from neural network analysis. Furthermore, the LRP method is also applied to identify and interpret the spatial patterns enabling the MLP decision-making process in surface temperature anomalies predictions on decadal timescales~\citep{toms2021assessing}, Madden-Julian Oscillation phase identification~\citep{toms2021testing}, and annual-mean temperature and precipitation predictions~\citep{barnes2020indicator}. Additionally, LRP is employed to elucidate the predictive skills of ensemble MLP for tracking global heating with ocean regimes~\citep{sonnewald2021revealing}. The application of LRP provides a post-hoc assessment of how ensemble MLP adjusts at each location in the North Atlantic region. This approach allows for a nuanced understanding of the contributions of various features like wind stress curl, latitudinal and longitudinal gradients, to the neural network's decision-making process, thereby enhancing confidence in the model's predictions and its application to unseen models or under different climate forcing scenarios.

\subsection{Other Explanation Methods}
Beyond the aforementioned three major categories of explanation techniques, there also exist some other post-hoc explanation methods such as activation maximization.
Activation maximization generates archetypal model inputs that highly activate certain layers like hidden neurons. This probes what features models have learned to detect. For example, activation maximization is used as a visualization technique to identify patterns that maximize specific activation functions in deep learning model for wind speed forecast~\citep{abdellaoui2020deep}. This post-hoc method focuses on finding new input data that maximizes the activation of a neuron, aiming to understand the model after training. The objective is to find input data that contribute the most to minimizing the error between the wind speed prediction and ground-truth data. For this purpose, the study defines a custom objective function as the inverse of the mean squared error, focusing on weather element forecasting as a regression problem. The interpreted results help further understand the most important features of weather forecasting in target cities.

Some other explainability tools have also been used to provide explanations for predictions made by machine learning models. For example, a study introduces a multi-variate wind speed forecasting model that leverages machine learning and a clustering-based multi-objective gravity search algorithm for improved accuracy and reliability~\citep{li2023wind}. Explainability in wind speed forecasting is implemented by employing post-hoc attribution analysis and visualization tools like partial dependence plots and individual conditional expectation plots. These tools help in understanding the model's interpretability, assessing the robustness and reliability of forecasted results, and explaining the relationship between features and forecast outcomes.

\section{Design Intrinsic Self-Explainable Model}\label{sec:Design Intrinsic Self-Explainable Model}
In this section, we introduce techniques to discuss more inherently interpretable models, with the aim of making the logic behind the predictions more transparent while still maintaining high accuracy. This is achieved by designing model architectures and components that are intuitive for humans to understand. However, perfectly interpretable models typically sacrifice some prediction performance. Thus, there is a trade-off between accuracy and interpretability that depends on the specific application. In the following, we summarize some of the key techniques to enhance inherent model transparency, including linear models, tree-based models, and attention mechanisms. We have also summarized these techniques in Table~\ref{tab:self-explain-applications}.

\begin{table}[t]
\small
\caption{An overview of intrinsic self-explainable models in weather and climate prediction.}
\begin{centering}
\begin{tabularx}{\textwidth}{m{4.5cm}m{5cm}X}
\toprule
\textbf{Self-explainable Models} & \textbf{Applcations} & \textbf{Prediction Tasks} \\
\toprule
\multirow{1}{=}{Linear Models} & 
 ~\citep{herman2018dendrology} & Extreme weather forecast \\
\hline
\multirow{3}{=}{Tree-based Models} & ~\citep{mecikalski2015probabilistic} & Convective initiation prediction \\
 & ~\citep{loken2022comparing} & Severe weather hazards prediction \\
 & ~\citep{zhang2019prediction} & Tropical cyclone genesis prediction \\
\hline
\multirow{2}{=}{Attention-based Explainable Neural Networks} & 
 ~\citep{tekin2021spatio} &  High-resolution temperature forecasting \\
 & ~\citep{suleman2022short} &  Short-term temperature forecasting \\
\hline
\end{tabularx}
\label{tab:self-explain-applications}
\end{centering}
\end{table}

\subsection{Linear Models}

Linear models such as linear regression and logistic regression (LR)  have coefficients that directly indicate the strength of association between the input predictors and the prediction. This inherent linear additivity makes the prediction explanation clear. However, linearity is often a simplifying assumption that can significantly reduce prediction accuracy. In general, linear models trade some precision for interpretability. To solve this, linear models can be combined with other machine learning models to obtain a forecast result with ideal accuracy and interpretability. For example, one study investigates the application of LR to extreme weather forecast, to understand how it makes predictions~\citep{herman2018dendrology}. The work performs a principal component analysis (PCA) before performing the prediction task, and the extracted principal components are provided to the LR. The authors show how visualizing regression coefficients for LR provides insight into what dynamics and variables are the most predictive. For example, the models automatically learn that model precipitation forecasts are most predictive along the Pacific coast where large-scale dynamics dominate extreme events, while moisture and instability become more important in the central US where convection is key factor. Analyzing these models builds understanding of the captured relationships, reveals systematic model biases like displacing precipitation features, and helps guide meteorological forecasters in using and correcting the guidance.

\subsection{Tree Based Models}
Decision tree and ensemble of trees are another important family of inherently interpretable models.
Decision trees partition the data space into rectilinear regions, with splits chosen to maximize information gain. The tree structure shows how predictions are made based on input variables that meet certain conditions. Ensembles of decision trees, such as random forests, improve accuracy while retaining some model transparency through tools like variable importance scores. For example, one study explores the use of random forests method for improving convective initiation predictions~\citep{mecikalski2015probabilistic}. It discusses the methodology for calculating feature importance in random forests. This process involves replacing each predictor variable with a randomized resampling and measuring the impact of this replacement on the random forests trees' prediction accuracy. Variables that significantly degrade accuracy when randomized are deemed the most important. The most critical findings reveal that two measures each of Convective Inhibition (CIN) and Convective Available Potential Energy (CAPE) are the most important predictors. 
Another study focuses on the use of random forests for predicting severe weather hazards~\citep{loken2022comparing}. It compares two methods of creating random forests based prediction for next-day severe weather using simulated data from the High Resolution Ensemble Forecast (HREF) system.  In both models, storm variables are identified as the most crucial, followed by index and environment variables. Additionally, they use a Python module tree interpreter to evaluate variable importance and the relationships acquired by the random forests, and find that the model focuses on  different features when predicting different hazards in a physically meaningful way. In addition to random forests, other tree-based machine learning models such as AdaBoost are also used in weather forecast, like predicting the evolution of Mesoscale Convective Systems (MCS) into tropical cyclones ~\citep{zhang2019prediction}. The AdaBoost classifier is built on environmental predictors and MCS properties known to influence tropical cyclones genesis. The model is used to quantify the contribution of each critical predictor to these classifiers' performance, contributing to the uncovering of new aspects of the tropical cyclones genesis.

\subsection{Attention-based Interpretable Models}

Beyond linear and tree-based models, DNNs via attention mechanism is another important inherently interpretable model family.
Some work explores making complex neural networks more interpretable by incorporating transparent model components such as attention layers. Adding interpretability modules increases understandability without sacrificing too much prediction performance. For example, one study discusses an innovative deep learning architecture for high-resolution numerical weather forecasting~\citep{tekin2021spatio}. This architecture utilizes Convolutional Long Short-Term Memory (ConvLSTM) and CNN units, integrated with an encoder-decoder structure. The model's interpretability and performance are enhanced by integrating the attention and context matching mechanism. The experiments conducted on the ERA5 hourly dataset demonstrate significant improvements in capturing spatial and temporal correlations for temperature forecasting, outperforming baseline models like ConvLSTM and U-Net.
Another study presents a novel deep learning model, Spatial Feature Attention Long Short-Term Memory (SFA-LSTM), designed for accurate temperature forecasting~\citep{suleman2022short}. Using an encoder-decoder architecture with LSTM layers, this model efficiently captures both spatial and temporal relationships among various meteorological variables when applied to temperature forecast. The spatial feature attention mechanism within the model enhances interpretability, allowing it to accurately forecast temperature changes by understanding the mutual influence of input weather variables. The model's performance, verified through domain knowledge, shows high accuracy and interpretability, especially in predicting temperature changes in weather forecast.

\section{Research Challenges}
Despite the current research progress in applying explainability to meteorological applications, there are also some research challenges remaining.

\subsection{Mechanistic Interpretability}
Recent advances in explainability have significantly impacted meteorological research, offering new insights and predictive capabilities. However, a critical limitation in the current explainability research is the heavy reliance on feature attributions. These methods, while useful in identifying which features in the data are most influential in the model's predictions, often fall short in providing deeper understanding of the underlying mechanisms. This superficial level of interpretation can be especially problematic in meteorology, where the causal relationships and dynamic interactions are complex and critical for accurate predictions and understanding. 
Feature attributions can highlight influential factors but without elucidating how these factors interact and contribute to the overall system dynamics. For instance, a model might identify water vapor and vertical velocity as key features in predicting rainfall, but it does not explain the mechanistic relationship between these variables and how they lead to precipitation. This gap hinders the ability of meteorologists to fully trust and understand AI predictions, which is crucial for high-stakes decision-making and advancing scientific knowledge.

Mechanistic interpretability, on the other hand, aims to address these shortcomings by providing insights into the `why' and `how' behind AI predictions~\citep{olah2020zoom,conmy2023towards,zhao2024opening}. It seeks to uncover the underlying causal relationships and principles that drive the model's output, aligning more closely with scientific discovery and reasoning. In meteorology, this means not just identifying important features, but understanding how these features interact in atmospheric and related systems (e.g., ocean, land and biosphere) to produce specific weather events or patterns.
The need for mechanistic interpretability in meteorology is two-fold. Firstly, it improves the credibility and trustworthiness of AI models by aligning their functioning with known scientific principles and theories. Secondly, it contributes to scientific discovery by potentially uncovering new insights and relationships within meteorological data that are not apparent through traditional analysis.

\subsection{Interpretability Evaluation}
Despite advances in developing explanation techniques, evaluating their utility and faithfulness remains an open challenge, largely due to the unavailability of ground-truth explanations. Although some studies have assessed different explainability  methods applied to climate science, these researches are mainly based on existing benchmarks and evaluation methods not standardized for the characteristics of climate data~\citep{mamalakis2022investigating,bommer2023finding}. Without an objective standardized measure of the quality of the explanation, it is extremely difficult to rigorously assess different methods and ensure that they provide meaningful insights aligned with reality. Therefore, the development of standardized benchmarks and metrics is critical for systematically evaluating explanation fidelity. Possible solutions involve designing proxy ground truth and pseudo ground-truth for the interpreted results~\citep{li2023mathcal}.
This requires carefully designed datasets with known explanatory factors, allowing explanation accuracy to be measured against ground-truth explanations. However, crafting appropriate benchmarks is highly complex for meteorological data, given intricate dynamical interdependencies. Collaboration with domain experts is essential to validate that learned relationships are scientifically sound.

\subsection{Making Use of Interpretability}
\begin{figure}[t]
  \centering
  \includegraphics[width=0.98\textwidth]{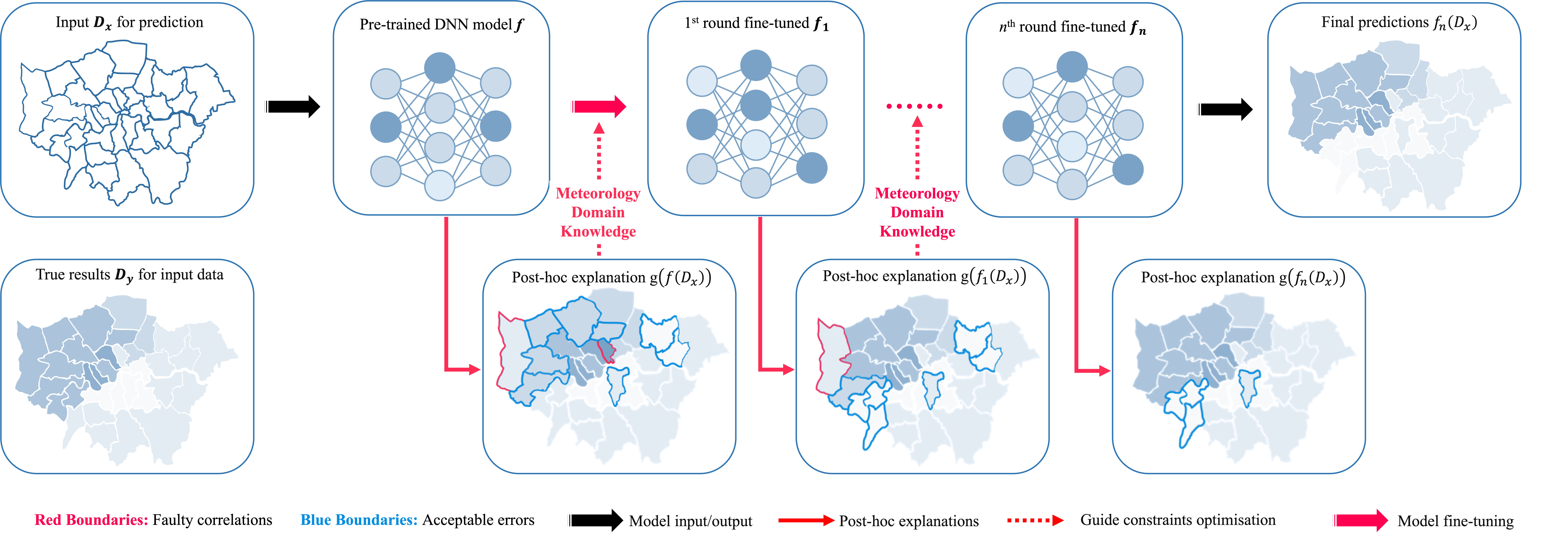}
  \caption{An example of using explainability  to guide iterative physical consistency improvement of a machine learning model (DNN model) until no faulty corrections (red boundaries) are detected.}
  \label{fig:make use of XAI}
\end{figure}

Explainable AI techniques provide valuable insights into the rationale behind the predictions of complex machine learning models. However, the full potential of interpretability lies in utilizing those insights to further improve model performance. The integration of explainability analysis into model development workflows can enable developers to continuously monitor feature attributions and model behaviors. This allows identifying areas where the model violates known constraints or relies on faulty correlations, guiding iterative refinement steps (Fig.~\ref{fig:make use of XAI}).

In particular, explainability can be used to analyze whether data-driven deep learning models make predictions that violate established physical principles or constraints. Since data-driven models are not based on encoding physics-based equations, they may learn spurious correlations that produce forecasts conflicting with the laws of atmospheric sciences. By attributing predictions to input features and visualizing the learned representations, developers can identify unreasonable logic and relationships in the model.
Once violations of physical consistency are detected through explanations, constraints can be introduced into model training and architectures. This includes adding loss penalty terms based on physical principles, using physics-infused neural networks, or representing conservation properties. Imposing appropriate physical constraints helps prevent models from making unreal predictions while still leveraging the flexibility of machine learning. This will eventually shift deep learning models closer to a hybrid data-driven and physics-based paradigm.

\subsection{Interpretability for Large Foundataion Models}
Recently, large foundation models have demonstrated impressive capabilities in various meteorological tasks such as weather forecasting, climate analysis, and understanding atmospheric processes~\citep{chen2023foundation}.
These models like  ClimaX~\citep{nguyen2023climax}, GraphCast~\citep{lam2022graphcast}, Fengwu~\citep{chen2023fengwu}, Fuxi~\citep{chen2023fuxi}, OceanGPT~\citep{bi2023oceangpt} leverage massive training datasets and billions of parameters to learn complex patterns and relationships. Nowadays, these large models are increasingly being deployed in weather and climate forecasting and decision support systems. However, their large scale and complexity introduce significant challenges for interpretability. 
Key challenges include faithfully localizing decision factors across billions of parameters, explaining how heterogeneous data modalities are integrated, and providing human-understandable explanations beyond low-level attributions. Potential solutions involve developing sparse attribution methods to isolate critical components, and using higher-level concept-based explanations based on atmospheric science. Besides, we can integrate causal reasoning techniques for mechanistic interpretations and create interactive interfaces for explanation at different granularities.

\section{Conclusions}
This survey presents a comprehensive review of interpretable machine learning techniques applied to weather and climate prediction. First, we provide an overview of how machine learning techniques are being applied to weather and climate prediction tasks. Then, we categorize the explainability methods into two main paradigms: post-hoc explanation approaches that interpret pre-trained models, and inherently interpretable model architectures designed from scratch. We analyze representative post-hoc techniques like SHAP, LRP, Grad-CAM, and LIME, highlighting how they uncover novel meteorological relationships captured by machine learning models. We also examine self-explainable model families such as linear models, tree ensembles, and attention-based neural networks that enhance transparency. Lastly, we have also summarized the open challenges around developing customized mechanistic interpretability methods to provide a more in-depth understanding of these methodology prediction models, evaluating explanation utility, leveraging insights obtained from explainability to improve the consistency of ML models with physical mechanisms, and developing effective explanation method for large foundation models.

\bibliography{tmlr}

\begin{thebibliography}{129}
\providecommand{\natexlab}[1]{#1}
\providecommand{\url}[1]{\texttt{#1}}
\expandafter\ifx\csname urlstyle\endcsname\relax
  \providecommand{\doi}[1]{doi: #1}\else
  \providecommand{\doi}{doi: \begingroup \urlstyle{rm}\Url}\fi

\bibitem[Abdellaoui \& Mehrkanoon(2020)Abdellaoui and Mehrkanoon]{abdellaoui2020deep}
Ismail~Alaoui Abdellaoui and Siamak Mehrkanoon.
\newblock Deep multi-stations weather forecasting: explainable recurrent convolutional neural networks.
\newblock \emph{arXiv preprint arXiv:2009.11239}, 2020.

\bibitem[Agrawal et~al.(2019)Agrawal, Barrington, Bromberg, Burge, Gazen, and Hickey]{agrawal2019machine}
Shreya Agrawal, Luke Barrington, Carla Bromberg, John Burge, Cenk Gazen, and Jason Hickey.
\newblock Machine learning for precipitation nowcasting from radar images.
\newblock \emph{arXiv preprint arXiv:1912.12132}, 2019.

\bibitem[Anderson et~al.(2009)Anderson, Hoar, Raeder, Liu, Collins, Torn, and Avellano]{anderson2009data}
Jeffrey Anderson, Tim Hoar, Kevin Raeder, Hui Liu, Nancy Collins, Ryan Torn, and Avelino Avellano.
\newblock The data assimilation research testbed: A community facility.
\newblock \emph{Bulletin of the American Meteorological Society}, 90\penalty0 (9):\penalty0 1283--1296, 2009.

\bibitem[Arcomano et~al.(2020)Arcomano, Szunyogh, Pathak, Wikner, Hunt, and Ott]{arcomano2020machine}
Troy Arcomano, Istvan Szunyogh, Jaideep Pathak, Alexander Wikner, Brian~R Hunt, and Edward Ott.
\newblock A machine learning-based global atmospheric forecast model.
\newblock \emph{Geophysical Research Letters}, 47\penalty0 (9):\penalty0 e2020GL087776, 2020.

\bibitem[Arcucci et~al.(2021)Arcucci, Zhu, Hu, and Guo]{arcucci2021deep}
Rossella Arcucci, Jiangcheng Zhu, Shuang Hu, and Yi-Ke Guo.
\newblock Deep data assimilation: integrating deep learning with data assimilation.
\newblock \emph{Applied Sciences}, 11\penalty0 (3):\penalty0 1114, 2021.

\bibitem[Arrieta et~al.(2020)Arrieta, D{\'\i}az-Rodr{\'\i}guez, Del~Ser, Bennetot, Tabik, Barbado, Garc{\'\i}a, Gil-L{\'o}pez, Molina, Benjamins, et~al.]{arrieta2020explainable}
Alejandro~Barredo Arrieta, Natalia D{\'\i}az-Rodr{\'\i}guez, Javier Del~Ser, Adrien Bennetot, Siham Tabik, Alberto Barbado, Salvador Garc{\'\i}a, Sergio Gil-L{\'o}pez, Daniel Molina, Richard Benjamins, et~al.
\newblock Explainable artificial intelligence (xai): Concepts, taxonomies, opportunities and challenges toward responsible ai.
\newblock \emph{Information fusion}, 58:\penalty0 82--115, 2020.

\bibitem[Ayzel et~al.(2020)Ayzel, Scheffer, and Heistermann]{ayzel2020rainnet}
Georgy Ayzel, Tobias Scheffer, and Maik Heistermann.
\newblock Rainnet v1. 0: a convolutional neural network for radar-based precipitation nowcasting.
\newblock \emph{Geoscientific Model Development}, 13\penalty0 (6):\penalty0 2631--2644, 2020.

\bibitem[Bach et~al.(2015)Bach, Binder, Montavon, Klauschen, M{\"u}ller, and Samek]{bach2015pixel}
Sebastian Bach, Alexander Binder, Gr{\'e}goire Montavon, Frederick Klauschen, Klaus-Robert M{\"u}ller, and Wojciech Samek.
\newblock On pixel-wise explanations for non-linear classifier decisions by layer-wise relevance propagation.
\newblock \emph{PloS one}, 10\penalty0 (7):\penalty0 e0130140, 2015.

\bibitem[Barnes et~al.(2020)Barnes, Toms, Hurrell, Ebert-Uphoff, Anderson, and Anderson]{barnes2020indicator}
Elizabeth~A Barnes, Benjamin Toms, James~W Hurrell, Imme Ebert-Uphoff, Chuck Anderson, and David Anderson.
\newblock Indicator patterns of forced change learned by an artificial neural network.
\newblock \emph{Journal of Advances in Modeling Earth Systems}, 12\penalty0 (9):\penalty0 e2020MS002195, 2020.

\bibitem[Bauer et~al.(2015)Bauer, Thorpe, and Brunet]{bauer2015quiet}
Peter Bauer, Alan Thorpe, and Gilbert Brunet.
\newblock The quiet revolution of numerical weather prediction.
\newblock \emph{Nature}, 525\penalty0 (7567):\penalty0 47--55, 2015.

\bibitem[Bi et~al.(2023{\natexlab{a}})Bi, Xie, Zhang, Chen, Gu, and Tian]{bi2023accurate}
Kaifeng Bi, Lingxi Xie, Hengheng Zhang, Xin Chen, Xiaotao Gu, and Qi~Tian.
\newblock Accurate medium-range global weather forecasting with 3d neural networks.
\newblock \emph{Nature}, 619\penalty0 (7970):\penalty0 533--538, 2023{\natexlab{a}}.

\bibitem[Bi et~al.(2023{\natexlab{b}})Bi, Zhang, Xue, Ou, Ji, Zheng, and Chen]{bi2023oceangpt}
Zhen Bi, Ningyu Zhang, Yida Xue, Yixin Ou, Daxiong Ji, Guozhou Zheng, and Huajun Chen.
\newblock Oceangpt: A large language model for ocean science tasks.
\newblock \emph{arXiv preprint arXiv:2310.02031}, 2023{\natexlab{b}}.

\bibitem[Bochenek \& Ustrnul(2022)Bochenek and Ustrnul]{bochenek2022machine}
Bogdan Bochenek and Zbigniew Ustrnul.
\newblock Machine learning in weather prediction and climate analyses—applications and perspectives.
\newblock \emph{Atmosphere}, 13\penalty0 (2):\penalty0 180, 2022.

\bibitem[Bodini et~al.(2020)Bodini, Lundquist, and Optis]{bodini2020can}
Nicola Bodini, Julie~K Lundquist, and Mike Optis.
\newblock Can machine learning improve the model representation of turbulent kinetic energy dissipation rate in the boundary layer for complex terrain?
\newblock \emph{Geoscientific Model Development}, 13\penalty0 (9):\penalty0 4271--4285, 2020.

\bibitem[Bommer et~al.(2023)Bommer, Kretschmer, Hedstr{\"o}m, Bareeva, and H{\"o}hne]{bommer2023finding}
Philine Bommer, Marlene Kretschmer, Anna Hedstr{\"o}m, Dilyara Bareeva, and Marina M-C H{\"o}hne.
\newblock Finding the right xai method--a guide for the evaluation and ranking of explainable ai methods in climate science.
\newblock \emph{arXiv preprint arXiv:2303.00652}, 2023.

\bibitem[Breiman(2001)]{breiman2001random}
Leo Breiman.
\newblock Random forests.
\newblock \emph{Machine learning}, 45:\penalty0 5--32, 2001.

\bibitem[Chen et~al.(2023{\natexlab{a}})Chen, Han, Gong, Bai, Ling, Luo, Chen, Ma, Zhang, Su, et~al.]{chen2023fengwu}
Kang Chen, Tao Han, Junchao Gong, Lei Bai, Fenghua Ling, Jing-Jia Luo, Xi~Chen, Leiming Ma, Tianning Zhang, Rui Su, et~al.
\newblock Fengwu: Pushing the skillful global medium-range weather forecast beyond 10 days lead.
\newblock \emph{arXiv preprint arXiv:2304.02948}, 2023{\natexlab{a}}.

\bibitem[Chen et~al.(2023{\natexlab{b}})Chen, Zhong, Zhang, Cheng, Xu, Qi, and Li]{chen2023fuxi}
Lei Chen, Xiaohui Zhong, Feng Zhang, Yuan Cheng, Yinghui Xu, Yuan Qi, and Hao Li.
\newblock Fuxi: A cascade machine learning forecasting system for 15-day global weather forecast.
\newblock \emph{arXiv preprint arXiv:2306.12873}, 2023{\natexlab{b}}.

\bibitem[Chen et~al.(2023{\natexlab{c}})Chen, Long, Jiang, Liu, and Zhang]{chen2023foundation}
Shengchao Chen, Guodong Long, Jing Jiang, Dikai Liu, and Chengqi Zhang.
\newblock Foundation models for weather and climate data understanding: A comprehensive survey.
\newblock \emph{arXiv preprint arXiv:2312.03014}, 2023{\natexlab{c}}.

\bibitem[Chkeir et~al.(2023)Chkeir, Anesiadou, Mascitelli, and Biondi]{chkeir2023nowcasting}
Sandy Chkeir, Aikaterini Anesiadou, Alessandra Mascitelli, and Riccardo Biondi.
\newblock Nowcasting extreme rain and extreme wind speed with machine learning techniques applied to different input datasets.
\newblock \emph{Atmospheric Research}, 282:\penalty0 106548, 2023.

\bibitem[Cintra \& de~Campos~Velho(2018)Cintra and de~Campos~Velho]{cintra2018data}
Rosangela~Saher Cintra and Haroldo~F de~Campos~Velho.
\newblock Data assimilation by artificial neural networks for an atmospheric general circulation model.
\newblock \emph{Advanced applications for artificial neural networks}, 265, 2018.

\bibitem[Conmy et~al.(2023)Conmy, Mavor-Parker, Lynch, Heimersheim, and Garriga-Alonso]{conmy2023towards}
Arthur Conmy, Augustine~N Mavor-Parker, Aengus Lynch, Stefan Heimersheim, and Adri{\`a} Garriga-Alonso.
\newblock Towards automated circuit discovery for mechanistic interpretability.
\newblock \emph{arXiv preprint arXiv:2304.14997}, 2023.

\bibitem[Daly(2006)]{daly2006guidelines}
Christopher Daly.
\newblock Guidelines for assessing the suitability of spatial climate data sets.
\newblock \emph{International Journal of Climatology: A Journal of the Royal Meteorological Society}, 26\penalty0 (6):\penalty0 707--721, 2006.

\bibitem[Deng et~al.(2021)Deng, Zou, Du, Chen, Feng, and Hu]{deng2021unified}
Huiqi Deng, Na~Zou, Mengnan Du, Weifu Chen, Guocan Feng, and Xia Hu.
\newblock A unified taylor framework for revisiting attribution methods.
\newblock In \emph{Proceedings of the AAAI Conference on Artificial Intelligence}, volume~35, pp.\  11462--11469, 2021.

\bibitem[Deng et~al.(2024)Deng, Zou, Du, Chen, Feng, Yang, Li, and Zhang]{deng2024unifying}
Huiqi Deng, Na~Zou, Mengnan Du, Weifu Chen, Guocan Feng, Ziwei Yang, Zheyang Li, and Quanshi Zhang.
\newblock Unifying fourteen post-hoc attribution methods with taylor interactions.
\newblock \emph{IEEE Transactions on Pattern Analysis and Machine Intelligence}, 2024.

\bibitem[Du et~al.(2019)Du, Liu, and Hu]{du2019techniques}
Mengnan Du, Ninghao Liu, and Xia Hu.
\newblock Techniques for interpretable machine learning.
\newblock \emph{Communications of the ACM}, 63\penalty0 (1):\penalty0 68--77, 2019.

\bibitem[Du et~al.(2020)Du, Bai, Tan, Xue, Samat, Xia, Li, Su, and Liu]{du2020advances}
Peijun Du, Xuyu Bai, Kun Tan, Zhaohui Xue, Alim Samat, Junshi Xia, Erzhu Li, Hongjun Su, and Wei Liu.
\newblock Advances of four machine learning methods for spatial data handling: A review.
\newblock \emph{Journal of Geovisualization and Spatial Analysis}, 4:\penalty0 1--25, 2020.

\bibitem[Dutta \& Pal(2022)Dutta and Pal]{dutta2022interpretation}
Debashree Dutta and Sankar~K Pal.
\newblock Interpretation of black box for short-term predictions of pre-monsoon cumulonimbus cloud events over kolkata.
\newblock \emph{Journal of Data, Information and Management}, 4\penalty0 (2):\penalty0 167--183, 2022.

\bibitem[Ebert-Uphoff \& Hilburn(2020)Ebert-Uphoff and Hilburn]{ebert2020evaluation}
Imme Ebert-Uphoff and Kyle Hilburn.
\newblock Evaluation, tuning and interpretation of neural networks for working with images in meteorological applications.
\newblock \emph{Bulletin of the American Meteorological Society}, pp.\  1--47, 2020.

\bibitem[Espeholt et~al.(2022)Espeholt, Agrawal, S{\o}nderby, Kumar, Heek, Bromberg, Gazen, Carver, Andrychowicz, Hickey, et~al.]{espeholt2022deep}
Lasse Espeholt, Shreya Agrawal, Casper S{\o}nderby, Manoj Kumar, Jonathan Heek, Carla Bromberg, Cenk Gazen, Rob Carver, Marcin Andrychowicz, Jason Hickey, et~al.
\newblock Deep learning for twelve hour precipitation forecasts.
\newblock \emph{Nature communications}, 13\penalty0 (1):\penalty0 1--10, 2022.

\bibitem[Foresti et~al.(2019)Foresti, Sideris, Nerini, Beusch, and Germann]{foresti2019using}
Loris Foresti, Ioannis~V Sideris, Daniele Nerini, Lea Beusch, and Urs Germann.
\newblock Using a 10-year radar archive for nowcasting precipitation growth and decay: A probabilistic machine learning approach.
\newblock \emph{Weather and Forecasting}, 34\penalty0 (5):\penalty0 1547--1569, 2019.

\bibitem[Fu et~al.(2021)Fu, Wang, Du, Liu, Hao, and Hu]{fu2021differentiated}
Weijie Fu, Meng Wang, Mengnan Du, Ninghao Liu, Shijie Hao, and Xia Hu.
\newblock Differentiated explanation of deep neural networks with skewed distributions.
\newblock \emph{IEEE Transactions on Pattern Analysis and Machine Intelligence}, 44\penalty0 (6):\penalty0 2909--2922, 2021.

\bibitem[Gagne~II et~al.(2019)Gagne~II, Haupt, Nychka, and Thompson]{gagne2019interpretable}
David~John Gagne~II, Sue~Ellen Haupt, Douglas~W Nychka, and Gregory Thompson.
\newblock Interpretable deep learning for spatial analysis of severe hailstorms.
\newblock \emph{Monthly Weather Review}, 147\penalty0 (8):\penalty0 2827--2845, 2019.

\bibitem[Gensini et~al.(2021)Gensini, Converse, Ashley, and Taszarek]{gensini2021machine}
Vittorio~A Gensini, Cody Converse, Walker~S Ashley, and Mateusz Taszarek.
\newblock Machine learning classification of significant tornadoes and hail in the united states using era5 proximity soundings.
\newblock \emph{Weather and Forecasting}, 36\penalty0 (6):\penalty0 2143--2160, 2021.

\bibitem[Ghada et~al.(2022)Ghada, Casellas, Herbinger, Garcia-Benad{\'\i}, Bothmann, Estrella, Bech, and Menzel]{ghada2022stratiform}
Wael Ghada, Enric Casellas, Julia Herbinger, Albert Garcia-Benad{\'\i}, Ludwig Bothmann, Nicole Estrella, Joan Bech, and Annette Menzel.
\newblock Stratiform and convective rain classification using machine learning models and micro rain radar.
\newblock \emph{Remote Sensing}, 14\penalty0 (18):\penalty0 4563, 2022.

\bibitem[Gibson et~al.(2021)Gibson, Chapman, Altinok, Delle~Monache, DeFlorio, and Waliser]{gibson2021training}
Peter~B Gibson, William~E Chapman, Alphan Altinok, Luca Delle~Monache, Michael~J DeFlorio, and Duane~E Waliser.
\newblock Training machine learning models on climate model output yields skillful interpretable seasonal precipitation forecasts.
\newblock \emph{Communications Earth \& Environment}, 2\penalty0 (1):\penalty0 159, 2021.

\bibitem[Giorgi \& Mearns(1991)Giorgi and Mearns]{giorgi1991approaches}
Filippo Giorgi and Linda~O Mearns.
\newblock Approaches to the simulation of regional climate change: a review.
\newblock \emph{Reviews of geophysics}, 29\penalty0 (2):\penalty0 191--216, 1991.

\bibitem[Gonz{\'a}lez-Abad et~al.(2023)Gonz{\'a}lez-Abad, Ba{\~n}o-Medina, and Guti{\'e}rrez]{gonzalez2023using}
Jose Gonz{\'a}lez-Abad, Jorge Ba{\~n}o-Medina, and Jos{\'e}~Manuel Guti{\'e}rrez.
\newblock Using explainability to inform statistical downscaling based on deep learning beyond standard validation approaches.
\newblock \emph{Journal of Advances in Modeling Earth Systems}, 15\penalty0 (11):\penalty0 e2023MS003641, 2023.

\bibitem[Griffin et~al.(2022)Griffin, Wimmers, and Velden]{griffin2022predicting}
Sarah~M Griffin, Anthony Wimmers, and Christopher~S Velden.
\newblock Predicting rapid intensification in north atlantic and eastern north pacific tropical cyclones using a convolutional neural network.
\newblock \emph{Weather and Forecasting}, 37\penalty0 (8):\penalty0 1333--1355, 2022.

\bibitem[Guo et~al.(2021)Guo, Lin, Wan, Li, and Cong]{guo2021learning}
Shengnan Guo, Youfang Lin, Huaiyu Wan, Xiucheng Li, and Gao Cong.
\newblock Learning dynamics and heterogeneity of spatial-temporal graph data for traffic forecasting.
\newblock \emph{IEEE Transactions on Knowledge and Data Engineering}, 34\penalty0 (11):\penalty0 5415--5428, 2021.

\bibitem[Gustafsson et~al.(2018)Gustafsson, Janji{\'c}, Schraff, Leuenberger, Weissmann, Reich, Brousseau, Montmerle, Wattrelot, Bu{\v{c}}{\'a}nek, et~al.]{gustafsson2018survey}
Nils Gustafsson, Tijana Janji{\'c}, Christoph Schraff, Daniel Leuenberger, Martin Weissmann, Hendrik Reich, Pierre Brousseau, Thibaut Montmerle, Eric Wattrelot, Anton{\'\i}n Bu{\v{c}}{\'a}nek, et~al.
\newblock Survey of data assimilation methods for convective-scale numerical weather prediction at operational centres.
\newblock \emph{Quarterly Journal of the Royal Meteorological Society}, 144\penalty0 (713):\penalty0 1218--1256, 2018.

\bibitem[Han et~al.(2021)Han, Chen, Chen, Chen, Zhang, Lu, Song, and Qin]{han2021deep}
Lei Han, Mingxuan Chen, Kangkai Chen, Haonan Chen, Yanbiao Zhang, Bing Lu, Linye Song, and Rui Qin.
\newblock A deep learning method for bias correction of ecmwf 24--240 h forecasts.
\newblock \emph{Advances in Atmospheric Sciences}, 38\penalty0 (9):\penalty0 1444--1459, 2021.

\bibitem[H{\"a}rter \& de~Campos~Velho(2012)H{\"a}rter and de~Campos~Velho]{harter2012data}
Fabr{\'\i}cio~P H{\"a}rter and Haroldo~Fraga de~Campos~Velho.
\newblock Data assimilation procedure by recurrent neural network.
\newblock \emph{Engineering Applications of Computational Fluid Mechanics}, 6\penalty0 (2), 2012.

\bibitem[He et~al.(2022)He, Li, Liu, Xu, Chen, Li, Zhang, Liu, Song, Xu, et~al.]{he2022improving}
Xinlei He, Yanping Li, Shaomin Liu, Tongren Xu, Fei Chen, Zhenhua Li, Zhe Zhang, Rui Liu, Lisheng Song, Ziwei Xu, et~al.
\newblock Improving predictions of land-atmosphere interactions based on a hybrid data assimilation and machine learning method.
\newblock \emph{Hydrology and Earth System Sciences Discussions}, 2022:\penalty0 1--33, 2022.

\bibitem[Herman \& Schumacher(2018)Herman and Schumacher]{herman2018dendrology}
Gregory~R Herman and Russ~S Schumacher.
\newblock “dendrology” in numerical weather prediction: What random forests and logistic regression tell us about forecasting extreme precipitation.
\newblock \emph{Monthly Weather Review}, 146\penalty0 (6):\penalty0 1785--1812, 2018.

\bibitem[Hersbach et~al.(2020)Hersbach, Bell, Berrisford, Hirahara, Hor{\'a}nyi, Mu{\~n}oz-Sabater, Nicolas, Peubey, Radu, Schepers, et~al.]{hersbach2020era5}
Hans Hersbach, Bill Bell, Paul Berrisford, Shoji Hirahara, Andr{\'a}s Hor{\'a}nyi, Joaqu{\'\i}n Mu{\~n}oz-Sabater, Julien Nicolas, Carole Peubey, Raluca Radu, Dinand Schepers, et~al.
\newblock The era5 global reanalysis.
\newblock \emph{Quarterly Journal of the Royal Meteorological Society}, 146\penalty0 (730):\penalty0 1999--2049, 2020.

\bibitem[Higa et~al.(2021)Higa, Tanahara, Adachi, Ishiki, Nakama, Yamada, Ito, Kitamoto, and Miyata]{higa2021domain}
Maiki Higa, Shinya Tanahara, Yoshitaka Adachi, Natsumi Ishiki, Shin Nakama, Hiroyuki Yamada, Kosuke Ito, Asanobu Kitamoto, and Ryota Miyata.
\newblock Domain knowledge integration into deep learning for typhoon intensity classification.
\newblock \emph{Scientific reports}, 11\penalty0 (1):\penalty0 12972, 2021.

\bibitem[Hilburn et~al.(2020)Hilburn, Ebert-Uphoff, and Miller]{hilburn2020development}
Kyle~A Hilburn, Imme Ebert-Uphoff, and Steven~D Miller.
\newblock Development and interpretation of a neural-network-based synthetic radar reflectivity estimator using goes-r satellite observations.
\newblock \emph{Journal of Applied Meteorology and Climatology}, 60\penalty0 (1):\penalty0 3--21, 2020.

\bibitem[Hu et~al.(2023)Hu, Ghazvinian, Chapman, Sengupta, Ralph, and Delle~Monache]{hu2023deep}
Weiming Hu, Mohammadvaghef Ghazvinian, William~E Chapman, Agniv Sengupta, Fred~Martin Ralph, and Luca Delle~Monache.
\newblock Deep learning forecast uncertainty for precipitation over the western united states.
\newblock \emph{Monthly Weather Review}, 151\penalty0 (6):\penalty0 1367--1385, 2023.

\bibitem[Kakkad et~al.(2023)Kakkad, Jannu, Sharma, Aggarwal, and Medya]{kakkad2023survey}
Jaykumar Kakkad, Jaspal Jannu, Kartik Sharma, Charu Aggarwal, and Sourav Medya.
\newblock A survey on explainability of graph neural networks.
\newblock \emph{arXiv preprint arXiv:2306.01958}, 2023.

\bibitem[Kalnay et~al.(2018)Kalnay, Kanamitsu, Kistler, Collins, Deaven, Gandin, Iredell, Saha, White, Woollen, et~al.]{kalnay2018ncep}
Eugenia Kalnay, Masao Kanamitsu, Robert Kistler, William Collins, Dennis Deaven, Lev Gandin, Mark Iredell, Suranjana Saha, Glenn White, John Woollen, et~al.
\newblock The ncep/ncar 40-year reanalysis project.
\newblock In \emph{Renewable energy}, pp.\  Vol1\_146--Vol1\_194. Routledge, 2018.

\bibitem[Kashinath et~al.(2021)Kashinath, Mustafa, Albert, Wu, Jiang, Esmaeilzadeh, Azizzadenesheli, Wang, Chattopadhyay, Singh, et~al.]{kashinath2021physics}
Karthik Kashinath, M~Mustafa, Adrian Albert, JL~Wu, C~Jiang, Soheil Esmaeilzadeh, Kamyar Azizzadenesheli, R~Wang, A~Chattopadhyay, A~Singh, et~al.
\newblock Physics-informed machine learning: case studies for weather and climate modelling.
\newblock \emph{Philosophical Transactions of the Royal Society A}, 379\penalty0 (2194):\penalty0 20200093, 2021.

\bibitem[Kochkov et~al.(2023)Kochkov, Yuval, Langmore, Norgaard, Smith, Mooers, Lottes, Rasp, D{\"u}ben, Kl{\"o}wer, et~al.]{kochkov2023neural}
Dmitrii Kochkov, Janni Yuval, Ian Langmore, Peter Norgaard, Jamie Smith, Griffin Mooers, James Lottes, Stephan Rasp, Peter D{\"u}ben, Milan Kl{\"o}wer, et~al.
\newblock Neural general circulation models.
\newblock \emph{arXiv preprint arXiv:2311.07222}, 2023.

\bibitem[Labe et~al.(2023)Labe, Johnson, and Delworth]{labe2023changes}
Zachary~M Labe, Nathaniel Johnson, and Thomas~L Delworth.
\newblock Changes in united states summer temperatures revealed by explainable neural networks.
\newblock \emph{Authorea Preprints}, 2023.

\bibitem[Lakshmanan et~al.(2015)Lakshmanan, Karstens, Krause, Elmore, Ryzhkov, and Berkseth]{lakshmanan2015polarimetric}
Valliappa Lakshmanan, Christopher Karstens, John Krause, Kim Elmore, Alexander Ryzhkov, and Samantha Berkseth.
\newblock Which polarimetric variables are important for weather/no-weather discrimination?
\newblock \emph{Journal of Atmospheric and Oceanic Technology}, 32\penalty0 (6):\penalty0 1209--1223, 2015.

\bibitem[Lam et~al.(2022)Lam, Sanchez-Gonzalez, Willson, Wirnsberger, Fortunato, Alet, Ravuri, Ewalds, Eaton-Rosen, Hu, et~al.]{lam2022graphcast}
Remi Lam, Alvaro Sanchez-Gonzalez, Matthew Willson, Peter Wirnsberger, Meire Fortunato, Ferran Alet, Suman Ravuri, Timo Ewalds, Zach Eaton-Rosen, Weihua Hu, et~al.
\newblock Graphcast: Learning skillful medium-range global weather forecasting.
\newblock \emph{arXiv preprint arXiv:2212.12794}, 2022.

\bibitem[Lam et~al.(2023)Lam, Sanchez-Gonzalez, Willson, Wirnsberger, Fortunato, Alet, Ravuri, Ewalds, Eaton-Rosen, Hu, et~al.]{lam2023learning}
Remi Lam, Alvaro Sanchez-Gonzalez, Matthew Willson, Peter Wirnsberger, Meire Fortunato, Ferran Alet, Suman Ravuri, Timo Ewalds, Zach Eaton-Rosen, Weihua Hu, et~al.
\newblock Learning skillful medium-range global weather forecasting.
\newblock \emph{Science}, 382\penalty0 (6677):\penalty0 1416--1421, 2023.

\bibitem[Lee et~al.(2020)Lee, Yoon, Jung, Chang, Kim, et~al.]{lee2020machine}
Seung~Hoon Lee, Yeon~Ah Yoon, Jin~Hyeong Jung, Tai-Woo Chang, Yong~Soo Kim, et~al.
\newblock A machine learning model for predicting silica concentrations through time series analysis of mining data.
\newblock \emph{Journal of Korean Society for Quality Management}, 48\penalty0 (3):\penalty0 511--520, 2020.

\bibitem[Legler \& Janji{\'c}(2022)Legler and Janji{\'c}]{legler2022combining}
Stefanie Legler and Tijana Janji{\'c}.
\newblock Combining data assimilation and machine learning to estimate parameters of a convective-scale model.
\newblock \emph{Quarterly Journal of the Royal Meteorological Society}, 148\penalty0 (743):\penalty0 860--874, 2022.

\bibitem[Leinonen et~al.(2020)Leinonen, Nerini, and Berne]{leinonen2020stochastic}
Jussi Leinonen, Daniele Nerini, and Alexis Berne.
\newblock Stochastic super-resolution for downscaling time-evolving atmospheric fields with a generative adversarial network.
\newblock \emph{IEEE Transactions on Geoscience and Remote Sensing}, 59\penalty0 (9):\penalty0 7211--7223, 2020.

\bibitem[Leinonen et~al.(2023)Leinonen, Hamann, Sideris, and Germann]{leinonen2023thunderstorm}
Jussi Leinonen, Ulrich Hamann, Ioannis~V Sideris, and Urs Germann.
\newblock Thunderstorm nowcasting with deep learning: A multi-hazard data fusion model.
\newblock \emph{Geophysical Research Letters}, 50\penalty0 (8):\penalty0 e2022GL101626, 2023.

\bibitem[Li et~al.(2023{\natexlab{a}})Li, Yang, He, Guo, Zhang, and Huang]{li2023wind}
Min Li, Yi~Yang, Zhaoshuang He, Xinbo Guo, Ruisheng Zhang, and Bingqing Huang.
\newblock A wind speed forecasting model based on multi-objective algorithm and interpretability learning.
\newblock \emph{Energy}, 269:\penalty0 126778, 2023{\natexlab{a}}.

\bibitem[Li et~al.(2023{\natexlab{b}})Li, Du, Chen, Chai, Lakkaraju, and Xiong]{li2023mathcal}
Xuhong Li, Mengnan Du, Jiamin Chen, Yekun Chai, Himabindu Lakkaraju, and Haoyi Xiong.
\newblock M4: A unified xai benchmark for faithfulness evaluation of feature attribution methods across metrics, modalities and models.
\newblock In \emph{Thirty-seventh Conference on Neural Information Processing Systems Datasets and Benchmarks Track}, 2023{\natexlab{b}}.

\bibitem[Li et~al.(2023{\natexlab{c}})Li, Liu, Shi, Chen, Zeng, Huo, and Fan]{li2023probabilistic}
Yang Li, Yubao Liu, Yueqin Shi, Baojun Chen, Fanhui Zeng, Zhaoyang Huo, and Hang Fan.
\newblock Probabilistic convective initiation nowcasting using himawari-8 ahi with explainable deep learning models.
\newblock \emph{Monthly Weather Review}, 2023{\natexlab{c}}.

\bibitem[Liu et~al.(2023)Liu, Lou, Yan, Qi, Jin, Yu, Yang, Zhao, and Xia]{liu2023deep}
Qi~Liu, Xiao Lou, Zhongwei Yan, Yajie Qi, Yuchao Jin, Shuang Yu, Xiaoliang Yang, Deming Zhao, and Jiangjiang Xia.
\newblock Deep-learning post-processing of short-term station precipitation based on nwp forecasts.
\newblock \emph{Atmospheric Research}, 295:\penalty0 107032, 2023.

\bibitem[Loken et~al.(2022)Loken, Clark, and McGovern]{loken2022comparing}
Eric~D Loken, Adam~J Clark, and Amy McGovern.
\newblock Comparing and interpreting differently designed random forests for next-day severe weather hazard prediction.
\newblock \emph{Weather and Forecasting}, 37\penalty0 (6):\penalty0 871--899, 2022.

\bibitem[Lu et~al.(2021)Lu, Ding, Yan, and Guo]{lu2021regional}
Zhiying Lu, Xudong Ding, Qin Yan, and Jianlin Guo.
\newblock Regional forecast of heavy precipitation and interpretability based on td-vae.
\newblock In \emph{2021 40th Chinese Control Conference (CCC)}, pp.\  7260--7265. IEEE, 2021.

\bibitem[Lundberg \& Lee(2017)Lundberg and Lee]{lundberg2017unified}
Scott~M Lundberg and Su-In Lee.
\newblock A unified approach to interpreting model predictions.
\newblock \emph{Advances in neural information processing systems}, 30, 2017.

\bibitem[Ma et~al.(2024)Ma, Liu, Dong, Chen, and Cai]{ma2024statistical}
Xingxing Ma, Hongnian Liu, Qiushi Dong, Qizhi Chen, and Ninghao Cai.
\newblock Statistical post-processing of multiple meteorological elements using the multimodel integration embedded method.
\newblock \emph{Atmospheric Research}, 301:\penalty0 107269, 2024.

\bibitem[Mamalakis et~al.(2022)Mamalakis, Barnes, and Ebert-Uphoff]{mamalakis2022investigating}
Antonios Mamalakis, Elizabeth~A Barnes, and Imme Ebert-Uphoff.
\newblock Investigating the fidelity of explainable artificial intelligence methods for applications of convolutional neural networks in geoscience.
\newblock \emph{Artificial Intelligence for the Earth Systems}, 1\penalty0 (4):\penalty0 e220012, 2022.

\bibitem[McGovern et~al.(2019)McGovern, Lagerquist, Gagne, Jergensen, Elmore, Homeyer, and Smith]{mcgovern2019making}
Amy McGovern, Ryan Lagerquist, David~John Gagne, G~Eli Jergensen, Kimberly~L Elmore, Cameron~R Homeyer, and Travis Smith.
\newblock Making the black box more transparent: Understanding the physical implications of machine learning.
\newblock \emph{Bulletin of the American Meteorological Society}, 100\penalty0 (11):\penalty0 2175--2199, 2019.

\bibitem[Mecikalski et~al.(2015)Mecikalski, Williams, Jewett, Ahijevych, LeRoy, and Walker]{mecikalski2015probabilistic}
John~R Mecikalski, John~K Williams, Christopher~P Jewett, David Ahijevych, Anita LeRoy, and John~R Walker.
\newblock Probabilistic 0--1-h convective initiation nowcasts that combine geostationary satellite observations and numerical weather prediction model data.
\newblock \emph{Journal of Applied Meteorology and Climatology}, 54\penalty0 (5):\penalty0 1039--1059, 2015.

\bibitem[Molina et~al.(2021)Molina, Gagne, and Prein]{molina2021benchmark}
Maria~J Molina, David~John Gagne, and Andreas~F Prein.
\newblock A benchmark to test generalization capabilities of deep learning methods to classify severe convective storms in a changing climate.
\newblock \emph{Earth and Space Science}, 8\penalty0 (9):\penalty0 e2020EA001490, 2021.

\bibitem[Molteni et~al.(1996)Molteni, Buizza, Palmer, and Petroliagis]{molteni1996ecmwf}
Franco Molteni, Roberto Buizza, Tim~N Palmer, and Thomas Petroliagis.
\newblock The ecmwf ensemble prediction system: Methodology and validation.
\newblock \emph{Quarterly journal of the royal meteorological society}, 122\penalty0 (529):\penalty0 73--119, 1996.

\bibitem[Montavon et~al.(2017)Montavon, Lapuschkin, Binder, Samek, and M{\"u}ller]{montavon2017explaining}
Gr{\'e}goire Montavon, Sebastian Lapuschkin, Alexander Binder, Wojciech Samek, and Klaus-Robert M{\"u}ller.
\newblock Explaining nonlinear classification decisions with deep taylor decomposition.
\newblock \emph{Pattern recognition}, 65:\penalty0 211--222, 2017.

\bibitem[Murdoch et~al.(2019)Murdoch, Singh, Kumbier, Abbasi-Asl, and Yu]{murdoch2019interpretable}
W~James Murdoch, Chandan Singh, Karl Kumbier, Reza Abbasi-Asl, and Bin Yu.
\newblock Interpretable machine learning: definitions, methods, and applications.
\newblock \emph{arXiv preprint arXiv:1901.04592}, 2019.

\bibitem[Nguyen et~al.(2023)Nguyen, Brandstetter, Kapoor, Gupta, and Grover]{nguyen2023climax}
Tung Nguyen, Johannes Brandstetter, Ashish Kapoor, Jayesh~K Gupta, and Aditya Grover.
\newblock Climax: A foundation model for weather and climate.
\newblock \emph{arXiv preprint arXiv:2301.10343}, 2023.

\bibitem[Olah et~al.(2020)Olah, Cammarata, Schubert, Goh, Petrov, and Carter]{olah2020zoom}
Chris Olah, Nick Cammarata, Ludwig Schubert, Gabriel Goh, Michael Petrov, and Shan Carter.
\newblock Zoom in: An introduction to circuits.
\newblock \emph{Distill}, 5\penalty0 (3):\penalty0 e00024--001, 2020.

\bibitem[Pan et~al.(2019)Pan, Hsu, AghaKouchak, and Sorooshian]{pan2019improving}
Baoxiang Pan, Kuolin Hsu, Amir AghaKouchak, and Soroosh Sorooshian.
\newblock Improving precipitation estimation using convolutional neural network.
\newblock \emph{Water Resources Research}, 55\penalty0 (3):\penalty0 2301--2321, 2019.

\bibitem[Pathak et~al.(2022)Pathak, Subramanian, Harrington, Raja, Chattopadhyay, Mardani, Kurth, Hall, Li, Azizzadenesheli, et~al.]{pathak2022fourcastnet}
Jaideep Pathak, Shashank Subramanian, Peter Harrington, Sanjeev Raja, Ashesh Chattopadhyay, Morteza Mardani, Thorsten Kurth, David Hall, Zongyi Li, Kamyar Azizzadenesheli, et~al.
\newblock Fourcastnet: A global data-driven high-resolution weather model using adaptive fourier neural operators.
\newblock \emph{arXiv preprint arXiv:2202.11214}, 2022.

\bibitem[Price et~al.(2023)Price, Sanchez-Gonzalez, Alet, Ewalds, El-Kadi, Stott, Mohamed, Battaglia, Lam, and Willson]{price2023gencast}
Ilan Price, Alvaro Sanchez-Gonzalez, Ferran Alet, Timo Ewalds, Andrew El-Kadi, Jacklynn Stott, Shakir Mohamed, Peter Battaglia, Remi Lam, and Matthew Willson.
\newblock Gencast: Diffusion-based ensemble forecasting for medium-range weather.
\newblock \emph{arXiv preprint arXiv:2312.15796}, 2023.

\bibitem[Qian \& Jia(2023)Qian and Jia]{qian2023seasonal}
QiFeng Qian and XiaoJing Jia.
\newblock Seasonal forecast of winter precipitation over china using machine learning models.
\newblock \emph{Atmospheric Research}, 294:\penalty0 106961, 2023.

\bibitem[Rajasekaran et~al.(2023)Rajasekaran, Sriram, Malini, and Sharma]{rajasekaran2023hybrid}
Umamaheswari Rajasekaran, GK~Sriram, A~Malini, and Vandana Sharma.
\newblock Hybrid explainable srnn-lstm architecture for irradiance, temperature and wind speed forecasting.
\newblock 2023.

\bibitem[Rampal et~al.(2022)Rampal, Gibson, Sood, Stuart, Fauchereau, Brandolino, Noll, and Meyers]{rampal2022high}
Neelesh Rampal, Peter~B Gibson, Abha Sood, Stephen Stuart, Nicolas~C Fauchereau, Chris Brandolino, Ben Noll, and Tristan Meyers.
\newblock High-resolution downscaling with interpretable deep learning: Rainfall extremes over new zealand.
\newblock \emph{Weather and Climate Extremes}, 38:\penalty0 100525, 2022.

\bibitem[Rasp \& Lerch(2018)Rasp and Lerch]{rasp2018neural}
Stephan Rasp and Sebastian Lerch.
\newblock Neural networks for postprocessing ensemble weather forecasts.
\newblock \emph{Monthly Weather Review}, 146\penalty0 (11):\penalty0 3885--3900, 2018.

\bibitem[Rasp et~al.(2018)Rasp, Pritchard, and Gentine]{rasp2018deep}
Stephan Rasp, Michael~S Pritchard, and Pierre Gentine.
\newblock Deep learning to represent subgrid processes in climate models.
\newblock \emph{Proceedings of the National Academy of Sciences}, 115\penalty0 (39):\penalty0 9684--9689, 2018.

\bibitem[Ren et~al.(2021)Ren, Li, Ren, Song, Xu, Deng, and Wang]{ren2021deep}
Xiaoli Ren, Xiaoyong Li, Kaijun Ren, Junqiang Song, Zichen Xu, Kefeng Deng, and Xiang Wang.
\newblock Deep learning-based weather prediction: a survey.
\newblock \emph{Big Data Research}, 23:\penalty0 100178, 2021.

\bibitem[Renault \& Mehrkanoon(2023)Renault and Mehrkanoon]{renault2023sar}
Mathieu Renault and Siamak Mehrkanoon.
\newblock Sar-unet: Small attention residual unet for explainable nowcasting tasks.
\newblock \emph{arXiv preprint arXiv:2303.06663}, 2023.

\bibitem[Retsch et~al.(2022)Retsch, Jakob, and Singh]{retsch2022identifying}
MH~Retsch, C~Jakob, and MS~Singh.
\newblock Identifying relations between deep convection and the large-scale atmosphere using explainable artificial intelligence.
\newblock \emph{Journal of Geophysical Research: Atmospheres}, 127\penalty0 (3):\penalty0 e2021JD035388, 2022.

\bibitem[Reulen \& Mehrkanoon(2024)Reulen and Mehrkanoon]{reulen2024ga}
Eloy Reulen and Siamak Mehrkanoon.
\newblock Ga-smaat-gnet: Generative adversarial small attention gnet for extreme precipitation nowcasting.
\newblock \emph{arXiv preprint arXiv:2401.09881}, 2024.

\bibitem[Ribeiro et~al.(2016{\natexlab{a}})Ribeiro, Singh, and Guestrin]{ribeiro2016model}
Marco~Tulio Ribeiro, Sameer Singh, and Carlos Guestrin.
\newblock Model-agnostic interpretability of machine learning.
\newblock \emph{arXiv preprint arXiv:1606.05386}, 2016{\natexlab{a}}.

\bibitem[Ribeiro et~al.(2016{\natexlab{b}})Ribeiro, Singh, and Guestrin]{ribeiro2016should}
Marco~Tulio Ribeiro, Sameer Singh, and Carlos Guestrin.
\newblock " why should i trust you?" explaining the predictions of any classifier.
\newblock In \emph{Proceedings of the 22nd ACM SIGKDD international conference on knowledge discovery and data mining}, pp.\  1135--1144, 2016{\natexlab{b}}.

\bibitem[Richardson(1922)]{richardson1922weather}
Lewis~F Richardson.
\newblock \emph{Weather prediction by numerical process}.
\newblock University Press, 1922.

\bibitem[Sachindra et~al.(2018)Sachindra, Ahmed, Rashid, Shahid, and Perera]{sachindra2018statistical}
DA~Sachindra, Khandakar Ahmed, Md~Mamunur Rashid, S~Shahid, and BJC Perera.
\newblock Statistical downscaling of precipitation using machine learning techniques.
\newblock \emph{Atmospheric research}, 212:\penalty0 240--258, 2018.

\bibitem[Scher \& Messori(2018)Scher and Messori]{scher2018predicting}
Sebastian Scher and Gabriele Messori.
\newblock Predicting weather forecast uncertainty with machine learning.
\newblock \emph{Quarterly Journal of the Royal Meteorological Society}, 144\penalty0 (717):\penalty0 2830--2841, 2018.

\bibitem[Seifert \& Rasp(2020)Seifert and Rasp]{seifert2020potential}
Axel Seifert and Stephan Rasp.
\newblock Potential and limitations of machine learning for modeling warm-rain cloud microphysical processes.
\newblock \emph{Journal of Advances in Modeling Earth Systems}, 12\penalty0 (12):\penalty0 e2020MS002301, 2020.

\bibitem[Selvaraju et~al.(2017)Selvaraju, Cogswell, Das, Vedantam, Parikh, and Batra]{selvaraju2017grad}
Ramprasaath~R Selvaraju, Michael Cogswell, Abhishek Das, Ramakrishna Vedantam, Devi Parikh, and Dhruv Batra.
\newblock Grad-cam: Visual explanations from deep networks via gradient-based localization.
\newblock In \emph{Proceedings of the IEEE international conference on computer vision}, pp.\  618--626, 2017.

\bibitem[Shi et~al.(2015)Shi, Chen, Wang, Yeung, Wong, and Woo]{shi2015convolutional}
Xingjian Shi, Zhourong Chen, Hao Wang, Dit-Yan Yeung, Wai-Kin Wong, and Wang-chun Woo.
\newblock Convolutional lstm network: A machine learning approach for precipitation nowcasting.
\newblock \emph{Advances in neural information processing systems}, 28, 2015.

\bibitem[Shield \& Houston(2022)Shield and Houston]{shield2022diagnosing}
Stephen~A Shield and Adam~L Houston.
\newblock Diagnosing supercell environments: A machine learning approach.
\newblock \emph{Weather and Forecasting}, 37\penalty0 (5):\penalty0 771--785, 2022.

\bibitem[Silva et~al.(2022)Silva, Keller, and Hardin]{silva2022using}
Sam~J Silva, Christoph~A Keller, and Joseph Hardin.
\newblock Using an explainable machine learning approach to characterize earth system model errors: Application of shap analysis to modeling lightning flash occurrence.
\newblock \emph{Journal of Advances in Modeling Earth Systems}, 14\penalty0 (4):\penalty0 e2021MS002881, 2022.

\bibitem[Sonnewald \& Lguensat(2021)Sonnewald and Lguensat]{sonnewald2021revealing}
Maike Sonnewald and Redouane Lguensat.
\newblock Revealing the impact of global heating on north atlantic circulation using transparent machine learning.
\newblock \emph{Journal of Advances in Modeling Earth Systems}, 13\penalty0 (8):\penalty0 e2021MS002496, 2021.

\bibitem[Suleman \& Shridevi(2022)Suleman and Shridevi]{suleman2022short}
Masooma Ali~Raza Suleman and S~Shridevi.
\newblock Short-term weather forecasting using spatial feature attention based lstm model.
\newblock \emph{IEEE Access}, 10:\penalty0 82456--82468, 2022.

\bibitem[Sundararajan et~al.(2017)Sundararajan, Taly, and Yan]{sundararajan2017axiomatic}
Mukund Sundararajan, Ankur Taly, and Qiqi Yan.
\newblock Axiomatic attribution for deep networks.
\newblock In \emph{International conference on machine learning}, pp.\  3319--3328. PMLR, 2017.

\bibitem[Tekin et~al.(2021)Tekin, Karaahmetoglu, Ilhan, Balaban, and Kozat]{tekin2021spatio}
Selim~Furkan Tekin, Oguzhan Karaahmetoglu, Fatih Ilhan, Ismail Balaban, and Suleyman~Serdar Kozat.
\newblock Spatio-temporal weather forecasting and attention mechanism on convolutional lstms.
\newblock \emph{arXiv preprint arXiv:2102.00696}, 4, 2021.

\bibitem[Thanh~Trieu et~al.(2021)Thanh~Trieu, Pottier, Rodin, and Xuan~Huynh]{thanh2021interpretable}
Ngoan Thanh~Trieu, Bernard Pottier, Vincent Rodin, and Hiep Xuan~Huynh.
\newblock Interpretable machine learning for meteorological data.
\newblock In \emph{2021 The 5th International Conference on Machine Learning and Soft Computing}, pp.\  11--17, 2021.

\bibitem[Toms et~al.(2020)Toms, Barnes, and Ebert-Uphoff]{toms2020physically}
Benjamin~A Toms, Elizabeth~A Barnes, and Imme Ebert-Uphoff.
\newblock Physically interpretable neural networks for the geosciences: Applications to earth system variability.
\newblock \emph{Journal of Advances in Modeling Earth Systems}, 12\penalty0 (9):\penalty0 e2019MS002002, 2020.

\bibitem[Toms et~al.(2021{\natexlab{a}})Toms, Barnes, and Hurrell]{toms2021assessing}
Benjamin~A Toms, Elizabeth~A Barnes, and James~W Hurrell.
\newblock Assessing decadal predictability in an earth-system model using explainable neural networks.
\newblock \emph{Geophysical Research Letters}, 48\penalty0 (12):\penalty0 e2021GL093842, 2021{\natexlab{a}}.

\bibitem[Toms et~al.(2021{\natexlab{b}})Toms, Kashinath, Yang, et~al.]{toms2021testing}
Benjamin~A Toms, Karthik Kashinath, Da~Yang, et~al.
\newblock Testing the reliability of interpretable neural networks in geoscience using the madden--julian oscillation.
\newblock \emph{Geoscientific Model Development}, 14\penalty0 (7):\penalty0 4495--4508, 2021{\natexlab{b}}.

\bibitem[Vald{\'e}s \& Pou(2021)Vald{\'e}s and Pou]{valdes2021machine}
Julio~J Vald{\'e}s and Antonio Pou.
\newblock A machine learning-explainable ai approach to tropospheric dynamics analysis using water vapor meteosat images.
\newblock In \emph{2021 IEEE Symposium Series on Computational Intelligence (SSCI)}, pp.\  1--8. IEEE, 2021.

\bibitem[Wang \& Li(2023)Wang and Li]{wang2023deep}
Chong Wang and Xiaofeng Li.
\newblock A deep learning model for estimating tropical cyclone wind radius from geostationary satellite infrared imagery.
\newblock \emph{Monthly Weather Review}, 151\penalty0 (2):\penalty0 403--417, 2023.

\bibitem[Wang et~al.(2020)Wang, Wang, Du, Yang, Zhang, Ding, Mardziel, and Hu]{wang2020score}
Haofan Wang, Zifan Wang, Mengnan Du, Fan Yang, Zijian Zhang, Sirui Ding, Piotr Mardziel, and Xia Hu.
\newblock Score-cam: Score-weighted visual explanations for convolutional neural networks.
\newblock In \emph{Proceedings of the IEEE/CVF Conference on Computer Vision and Pattern Recognition Workshops}, 2020.

\bibitem[Wang(2014)]{wang2014meteoinfo}
Yaqiang Wang.
\newblock Meteoinfo: Gis software for meteorological data visualization and analysis.
\newblock \emph{Meteorological Applications}, 21\penalty0 (2):\penalty0 360--368, 2014.

\bibitem[Wang(2019)]{wang2019open}
YaQiang Wang.
\newblock An open source software suite for multi-dimensional meteorological data computation and visualisation.
\newblock \emph{J. Open Res. Softw}, 7\penalty0 (1):\penalty0 21, 2019.

\bibitem[Wang et~al.(2022)Wang, Shi, Lei, and Fung]{wang2022deep}
Yueya Wang, Xiaoming Shi, Lili Lei, and Jimmy Chi-Hung Fung.
\newblock Deep learning augmented data assimilation: Reconstructing missing information with convolutional autoencoders.
\newblock \emph{Monthly Weather Review}, 150\penalty0 (8):\penalty0 1977--1991, 2022.

\bibitem[Watson(2019)]{watson2019applying}
Peter~AG Watson.
\newblock Applying machine learning to improve simulations of a chaotic dynamical system using empirical error correction.
\newblock \emph{Journal of Advances in Modeling Earth Systems}, 11\penalty0 (5):\penalty0 1402--1417, 2019.

\bibitem[Weyn et~al.(2019)Weyn, Durran, and Caruana]{weyn2019can}
Jonathan~A Weyn, Dale~R Durran, and Rich Caruana.
\newblock Can machines learn to predict weather? using deep learning to predict gridded 500-hpa geopotential height from historical weather data.
\newblock \emph{Journal of Advances in Modeling Earth Systems}, 11\penalty0 (8):\penalty0 2680--2693, 2019.

\bibitem[Weyn et~al.(2020)Weyn, Durran, and Caruana]{weyn2020improving}
Jonathan~A Weyn, Dale~R Durran, and Rich Caruana.
\newblock Improving data-driven global weather prediction using deep convolutional neural networks on a cubed sphere.
\newblock \emph{Journal of Advances in Modeling Earth Systems}, 12\penalty0 (9):\penalty0 e2020MS002109, 2020.

\bibitem[Wu et~al.(2021)Wu, Chang, Yuan, Sun, Zhang, Arcucci, and Guo]{wu2021fast}
Pin Wu, Xuting Chang, Wenyan Yuan, Junwu Sun, Wenjie Zhang, Rossella Arcucci, and Yike Guo.
\newblock Fast data assimilation (fda): Data assimilation by machine learning for faster optimize model state.
\newblock \emph{Journal of Computational Science}, 51:\penalty0 101323, 2021.

\bibitem[Xiong et~al.(2024)Xiong, Zhang, Chen, Sun, Li, Sun, Du, et~al.]{xiong2024towards}
Haoyi Xiong, Xiaofei Zhang, Jiamin Chen, Xinhao Sun, Yuchen Li, Zeyi Sun, Mengnan Du, et~al.
\newblock Towards explainable artificial intelligence (xai): A data mining perspective.
\newblock \emph{arXiv preprint arXiv:2401.04374}, 2024.

\bibitem[Yang et~al.(2022)Yang, Mu, Wang, and Wang]{yang2022hourly}
Ruyi Yang, Jianli Mu, Shudong Wang, and Lijuan Wang.
\newblock Hourly rolling correction of precipitation forecast via convolutional and long short-term memory networks.
\newblock \emph{Atmospheric Science Letters}, 23\penalty0 (10):\penalty0 e1100, 2022.

\bibitem[Yu \& Yang(2023)Yu and Yang]{yu2023temporal}
Tingzhao Yu and Ruyi Yang.
\newblock Temporal dynamic network with learnable coupled adjacent matrix for wind forecasting.
\newblock \emph{IEEE Geoscience and Remote Sensing Letters}, 2023.

\bibitem[Yu et~al.(2021)Yu, Kuang, and Yang]{yu2021atmconvgru}
Tingzhao Yu, Qiuming Kuang, and Ruyi Yang.
\newblock Atmconvgru for weather forecasting.
\newblock \emph{IEEE Geoscience and Remote Sensing Letters}, 19:\penalty0 1--5, 2021.

\bibitem[Yu et~al.(2022)Yu, Yang, Huang, Gao, and Kuang]{yu2022terrain}
Tingzhao Yu, Ruyi Yang, Yan Huang, Jinbing Gao, and Qiuming Kuang.
\newblock Terrain-guided flatten memory network for deep spatial wind downscaling.
\newblock \emph{IEEE Journal of Selected Topics in Applied Earth Observations and Remote Sensing}, 15:\penalty0 9468--9481, 2022.

\bibitem[Zhang et~al.(2020)Zhang, Zeng, Wang, Ma, and Chu]{zhang2020correction}
Chang-Jiang Zhang, Jing Zeng, Hui-Yuan Wang, Lei-Ming Ma, and Hai Chu.
\newblock Correction model for rainfall forecasts using the lstm with multiple meteorological factors.
\newblock \emph{Meteorological Applications}, 27\penalty0 (1):\penalty0 e1852, 2020.

\bibitem[Zhang et~al.(2019)Zhang, Lin, Lin, Zhang, Yu, Cao, and Xue]{zhang2019prediction}
Tao Zhang, Wuyin Lin, Yanluan Lin, Minghua Zhang, Haiyang Yu, Kathy Cao, and Wei Xue.
\newblock Prediction of tropical cyclone genesis from mesoscale convective systems using machine learning.
\newblock \emph{Weather and Forecasting}, 34\penalty0 (4):\penalty0 1035--1049, 2019.

\bibitem[Zhang et~al.(2023)Zhang, Long, Chen, Xing, Jin, Jordan, and Wang]{zhang2023skilful}
Yuchen Zhang, Mingsheng Long, Kaiyuan Chen, Lanxiang Xing, Ronghua Jin, Michael~I Jordan, and Jianmin Wang.
\newblock Skilful nowcasting of extreme precipitation with nowcastnet.
\newblock \emph{Nature}, 619\penalty0 (7970):\penalty0 526--532, 2023.

\bibitem[Zhao et~al.(2024)Zhao, Yang, Lakkaraju, and Du]{zhao2024opening}
Haiyan Zhao, Fan Yang, Himabindu Lakkaraju, and Mengnan Du.
\newblock Opening the black box of large language models: Two views on holistic interpretability.
\newblock \emph{arXiv preprint arXiv:2402.10688}, 2024.

\bibitem[Zhou et~al.(2020)Zhou, Zheng, Dong, and Wang]{zhou2020deep}
Kanghui Zhou, Yongguang Zheng, Wansheng Dong, and Tingbo Wang.
\newblock A deep learning network for cloud-to-ground lightning nowcasting with multisource data.
\newblock \emph{Journal of Atmospheric and Oceanic Technology}, 37\penalty0 (5):\penalty0 927--942, 2020.

\bibitem[Zhuo \& Tan(2021)Zhuo and Tan]{zhuo2021physics}
Jing-Yi Zhuo and Zhe-Min Tan.
\newblock Physics-augmented deep learning to improve tropical cyclone intensity and size estimation from satellite imagery.
\newblock \emph{Monthly Weather Review}, 149\penalty0 (7):\penalty0 2097--2113, 2021.

\end{thebibliography}
\bibliographystyle{tmlr}

\end{document}